\documentclass[aps,onecolumn,nofootinbib,preprintnumbers,superscriptaddress]{revtex4}

\usepackage{graphicx}
\usepackage{amssymb}
\usepackage{amsmath,bm}
\usepackage{amsfonts}         
\usepackage{fancybox}   
\usepackage{yfonts} 
\usepackage{epstopdf}
\usepackage{epsfig}
\usepackage[pdftex,colorlinks,citecolor=blue,linkcolor=red,urlcolor=blue]{hyperref} 
\usepackage{comment}

\newcommand{\be}{\begin{equation}}
\newcommand{\ee}{\end{equation}}
\newcommand{\bea}{\begin{eqnarray}}
\newcommand{\eea}{\end{eqnarray}}
\newcommand{\figref}[1]{Fig.\,\ref{#1}}

\newcommand{\Img}{\mathop{\mathrm{Im}}}
\newcommand{\Rea}{\mathop{\mathrm{Re}}}

\newcommand{\AdS}{\mathop{\mathrm{AdS}}}
\newcommand{\Schr}{\mathop{\mathrm{Schr}}}
\newcommand{\Dm}{\mathop{\mathrm{D_m}}}
\newcommand{\Cv}{\mathop{\mathrm{C_v}}}
\newcommand{\Tr}{\mathop{\mathrm{Tr}}}

\numberwithin{equation}{section}

\begin{document}

\title{Schr\"odinger Fermi Liquids} %
\author{Juven Wang} 
\affiliation{Department of Physics and Center for Theoretical Physics, Massachusetts Institute of Technology, Cambridge, MA 02139, USA} 
\affiliation{Perimeter Institute for Theoretical Physics, Waterloo, ON, N2L 2Y5, Canada}

\preprint{MIT-CTP/4354}

\begin{abstract}\noindent

A class of strongly interacting many-body fermionic systems in $2+1$-dimensional  
non-relativistic conformal field theory is examined via the gauge-gravity duality correspondence. 
The $5$-dimensional charged black hole with asymptotic Schr\"odinger isometry in the bulk gravity side introduces parameters of background density and finite particle number into the boundary field theory. We 
propose the holographic dictionary,
and realize a quantum phase transition of this fermionic liquid with fixed particle number by tuning the background density $\beta$ at zero temperature. On the larger $\beta$ side, we find the signal of a sharp quasiparticle pole on the spectral function $A(k,\omega)$, indicating a well-defined Fermi surface. On the smaller $\beta$ side, we find only a hump with no sharp peak for $A(k,\omega)$, indicating the disappearance of Fermi surface. The dynamical exponent $z$ of quasiparticle dispersion 
goes from being Fermi-liquid-like $z\simeq1$ scaling at larger $\beta$ to a non-Fermi-liquid scaling $z\simeq 3/2$ at smaller $\beta$.
By comparing the structure of Green's function with Landau Fermi liquid theory and Senthil's scaling ansatz, we further investigate the behavior of this quantum phase transition.

\end{abstract}

\maketitle
\setcounter{footnote}{0} 

\tableofcontents


\section{Introduction}

Experimental physics on strongly interacting non-relativistic many-body bosonic and fermionic systems develops rapidly on account of the progress of controlling ultra cold atoms\cite{Giorgini:2008zz,Bloch:2008zzb,Zwierlein}.
The dynamical exponent $z$ of those microscopic states is usually $z \neq 1$, 
along with other well-known condensed matter systems 
such as non-Fermi liquids metals from heavy fermions and high $T_c$ superconductor(see \cite{Si:2011hh,McGreevy:2010zz} and reference therein), beyond the paradigm of Landau Fermi liquid theory. 
In this strongly coupling regime, the traditional perturbative field theory study has been challenged, while
holography, specifically Anti de Sitter space and conformal field theory(AdS/CFT) correspondence\cite{Maldacena:1997re,Gubser:1998bc,Witten:1998qj}, becomes a powerful alternative. 
Holographic methods have shown some success in the study of certain strongly interacting fermion systems\cite{Lee:2008xf,Liu:2009dm,Cubrovic:2009ye,Faulkner:2009wj}, 
with the emergence of Fermi surface and non-Fermi liquid behavior.
(See \cite{Hartnoll:2009sz,McGreevy:2009xe,Hartnoll:2011fn} for reviews on the holography applied to condensed matter physics.)
However, there are at least two major shortcomings to bridge this success to ultra cold atomic systems.
On one hand, the dual field theories of these study are asymptotic conformal and relativistic, which are quite different from the non-relativistic nature of ultra cold atoms.
On the other hand, this AdS/CFT setting bears no parameters identifying the particle number, mass or density spectrum, because of these parameters are not good quantum numbers in the relativistic theory. Tuning physical parameters such as particle number or doping density 
becomes important in the absolute zero temperature, where purely quantum fluctuations can drive phase transitions, known as 
quantum phase transitions\cite{Sachdev:QPT}.\\

Substantial works in the literature had contributed to understand non-relativistic conformal field theory(NRCFT)(\cite{Mehen:1999nd,Nishida:2007pj,Nishida:2010tm} and reference therein), as a renormalization group(RG) fixed point of the non-relativistic systems.
And its gravity dual theory had been proposed, with solutions of zero temperature pure Schr\"odinger(Schr) geometry\cite{Son:2008ye,Balasubramanian:2008dm}, 
finite temperature black holes\cite{Adams:2008wt,Herzog:2008wg,Maldacena:2008wh}, 
and charged black holes\cite{Adams:2009dm,Imeroni:2009cs}.
There has been some pursuits on studying fermions in this asymptotic Schr\"odinger geometry\cite{Akhavan:2009ns,Leigh:2009ck,Hung:2010te}.
However, to our knowledge, the holographic study of Fermi surface from strongly interacting fermions with NRCFT background 
has not yet been explored in the literature. Our paper is aimed to bridge this gap and tackle the two aforementioned shortcomings.\\ 

Schr\"odinger black hole in the bulk gravity theory, on one hand, realizes an asymptotic NRCFT background with the dynamical exponent $z=2$ for the boundary field theory naturally.
On the other hand, Schr\"odinger black hole provides the particle number(or the mass operator\cite{Adams:2011kb}) $M$ from the gauge invariant $\xi$-momentum: $M=\ell-q A_\xi|_\partial$ \cite{Adams:2011kb}
and background density $\beta$ to the non-relativistic boundary field theory.
Our approach is similar to AdS/CFT set-up\cite{Liu:2009dm}, 
considering a Dirac fermion field in the probe limit under a charged black hole spacetime,
the Green's function can be read from the asymptotic behavior of Dirac fermion field in the UV of the bulk side.
Here we also propose the holographic dictionary of real-time retarded Green's function for fermions in Schr/NRCFT correspondence\cite{Mahajan}, analog to the work of \cite{Iqbal:2009fd} for AdS/CFT case.
For convenience, we name these classes of strongly interacting non-relativistic fermionic liquids under asymptotic NRCFT background as Schr\"odinger Fermi liquids.\\

The paper is organized as follows.
Firstly we discuss the charged Schr\"odinger black hole solution and its Dirac fermion equation of motion, to introduce our notations in Sec.\ref{sec2}.
We then provide our holographic dictionary\cite{Mahajan} analog to the setting of \cite{Liu:2009dm,Iqbal:2009fd} in Sec.\ref{sec3}.
In Sec.\ref{sec4}, we demonstrate the zero temperature nearly ground state of this fermion system shows a sharp quasiparticle peak in the spectral function - the evidence of Fermi surface, 
with a non-Fermi liquid dispersion relation. 
We compare our spectral functions $A(k,\omega)$ of Schr\"odinger Fermi liquids with Landau Fermi liquid theory and Senthil's scaling ansatz\cite{Senthil,senthil:0804}.
In Sec.\ref{sec5}, we show the evidence of a quantum phase transition,
by tuning the background density $\beta$ but fixing particle number at zero temperature. 
On the larger $\beta$ side, we find a well-defined Fermi surface.
On the smaller $\beta$ side, we find only a hump with no sharp peak for $A(k,\omega)$, indicating the disappearance of Fermi surface. 
The dynamical exponent $z$ of the quasiparticle dispersion goes from Fermi-liquid-like scaling $z=1$ at larger $\beta$ to larger $z(\simeq 3/2)$ non-Fermi liquid at smaller $\beta$.
Finally, we conclude with some remarks and open questions in Sec.\ref{sec6}.\\

Our program code for numerical computation is shared through this URL\cite{code}.

\section{Set-Up: Dirac Fermion Field in a Charged Schr\"odinger Black Hole} \label{sec2}
Based on the holography, a quantum field theory of finite charge density can be mapped to a charged black hole of a gravity theory\cite{Liu:2009dm}.
The U(1) charge of Schr\"odinger black hole induces finite charge density to the boundary field theory, meanwhile breaks the non-relativistic conformal invariance.
Thus, we only have `asymptotic' NRCFT.
Before discuss the details of bulk gravity theory,
it will be helpful to introduce generic labels for a large class of NRCFT(with charge and mass densities) we will study. 
We characterize our `asymptotic' NRCFT by five parameters, $(\Delta, M, \mu_Q, \beta, T)$\cite{Adams:2011kb}. 
Two parameters, the conformal dimension $\Delta$ and the mass operator $M=\ell-q A_\xi|_\partial$ from the gauge invariant $\xi$-momentum, 
define the boundary NRCFT 
in a universal sector. The U(1) charge chemical potential $\mu_Q$ and other relevant terms from the current $J_\mu$(such as charge density $\rho_{Q}$ and mass density $\rho_{M}$) in NRCFT is mapped to U(1) gauge field $A_\mu$ of the bulk gravity. %
Background density $\beta$ is introduced by Schr\"odinger black hole through Null Melvin Twist(or TsT transformation)\cite{Adams:2008wt,Herzog:2008wg,Maldacena:2008wh,Adams:2009dm,Imeroni:2009cs}. The physical way to interpret this $\beta$ could be the density of doping background, or an analog of interaction strength $t/U$ of Hubbard model\footnote{Indeed $\beta$ has dimension [length]$^{1}$, so the ``background density over area'' with correct dimension should be defined as $\beta^{-2}$. 
We simply name $\beta$ as background density for convenience.
}.
Temperature $T$ of the boundary theory is given by the black hole temperature $T_{BH}$.
Notably, the conformal dimension of NRCFT here depends on the mass operator $M$, which is quite different from CFT. 
More peculiarly, the conformal dimension for spinors has an extra $m\pm \frac{1}{2}$ split, as already been noticed in \cite{Akhavan:2009ns,Leigh:2009ck}.
In the following we denote dimension $d$ as the spatial dimension of $x_1,x_2,\dots, x_d$. 
Thus the bulk spacetime of Schr\"odinger asymptotic as Schr$_{d+3}$(distinguished from the AdS$_{d+2}$), 
where as the corresponding boundary theory of Schr$_{d+3}$ is NRCFT$_{d+1}$ (CFT$_{d+1}$ for AdS$_{d+2}$). 
We summarize the conformal dimensions of spin-0 boson\cite{Adams:2011kb} and spin-1/2 fermion operator(\cite{Akhavan:2009ns,Leigh:2009ck}) in the following table.
\begin{table}[h]
\begin{tabular}{c||c|c}
 asymptotics & $\AdS_{d+2}$ & $\Schr_{d+3}$\\ \hline
  scalar conformal dimension& $\Delta_{\pm}=\frac{d+1}{2} \pm \sqrt{(\frac{d+1}{2} )^2+m^2}$ & $\Delta_{\pm}=\frac{d+2}{2} \pm \sqrt{(\frac{d+2}{2} )^2+m^2+M^2}$   \\
spinor conformal dimension& $\Delta_{\pm}= \frac{d+1}{2}  \pm m$ & $\Delta_{\pm}=\frac{d+2}{2} \pm \sqrt{(m\pm \frac{1}{2})^2+M^2}$   \\
 \end{tabular}
\caption{conformal dimensions of CFT and NRCFT for spin-0 boson and spin-1/2 fermion}
 \label{conformal dim}
\end{table}

\subsection{Charged Schr\"odinger black hole}

We focus on $d=2$, 5-dimensional(5D) Schr\"odinger black hole Schr$_{5}$ in the bulk and 2+1D NRCFT$_{3}$ on the boundary.
Let us briefly go through our set-up for the charged $\Schr_5$. In string frame, the metric is,
\begin{equation}
\label{eq:mainmetrictauy}
ds_{Str}^2 = \frac{K r^2}{R^2} 
\left(
  -f d\tau^2 + dy^2 - \beta^2r^2f(d\tau + dy)^2
\right)
+
\frac{r^2}{R^2} (dx_1^2 + dx_2^2) + \frac{R^2}{r^2} \frac{dr^2}{f}.
\end{equation}
where R is curvature radius. By converting to light-cone like coordinates, 
$t = \beta(\tau + y), \quad \xi = \frac{1}{2\beta}(-\tau+y)$,
and switching to Einstein frame for the later use of holographic dictionary,

\begin{equation}
\label{eq:mainmetrictxi}
ds_{Ein}^2 = K^{-1/3} \Bigg(
 \frac{Kr^2}{R^2} 
\bigg(
  \Big(
    \frac{1-f}{4\beta^2} -r^2 f
  \Big) dt^2 
  + \beta^2(1-f) d\xi^2 
   +   (1+f) dtd\xi 
\bigg) \nonumber 
  + \frac{r^2}{R^2} (dx_1^2+dx_2^2) + \frac{R^2}{r^2} \frac{dr^2}{f} \Bigg)
\end{equation}

where
$f(r) = 1 + \frac{Q^2}{r^6} -
\left(r_0^4 + \frac{Q^2}{r_0^2}\right) \frac{1}{r^4},
K(r) = \frac{1}{1+\beta^2r^2(1-f(r))}.
$
The charged black hole supports a gauge field,
\begin{equation}
A = A_\tau \mathbf{d}\tau, \quad A_\tau = \frac{Q}{R^2r_0^2} \left(1-\frac{r_0^2}{r^2}\right),
\end{equation} 

Adopted in \cite{Adams:2011kb} notation for later use, $A=\frac{A_\tau}{2 \beta} dt-\beta A_\tau d\xi=A_t dt+A_\xi d \xi$, with\\
\be
A_t    = \mu_{Q} +\rho_{Q}\, r^{-2}\,, ~~~~~~~~~~
A_\xi = M_{o} + \rho_{M}\, r^{-2} 
\ee 
By holography\cite{Adams:2011kb}, $\mu_{Q}$ is identified as U(1) charge chemical potential, $\rho_{Q},\rho_{M}$ are the charge density and mass density, 
$M_{o}$ is related to the mass operator by $M=\ell-q A_\xi|_\partial=\ell-q M_{o}$ with $\ell$ as the $\xi$-momentum.
The temperature of the black hole is given by identifying the inverse of the near-horizon Euclidean periodicity of boundary time coordinate $t$,
\begin{equation}
T_{BH} = \frac{r_0}{\pi \beta R^2} 
\left(
1 - \frac{Q^2}{2r_0^6}
\right),
\end{equation}

Our interest of study is the boundary field theory at zero temperature, which corresponds to the extremal black hole with $Q=\sqrt{2} r_0^3$. 
This being said, all the numerical analysis contained in this paper pertains to zero temperature only.
At zero temperature, the charged black hole $\Schr_5$ horizons degenerate, meanwhile the near horizon geometry becomes 
AdS$_2\times \mathbb{R}^3$, with\footnote{see Appendix \ref{AdS2} for details}
$ds^2=- \epsilon^2 d\tilde{\tau}^2/R^2_{\AdS_2}+R^2_{\AdS_2}(d\epsilon^2/\epsilon^2)+r_0^2 d \vec{\tilde{x}}^2/R^2_{\\AdS_2}$, with $R_{\AdS_2}=\frac{(1+\beta^2 r_0^2)^{1/6}}{\sqrt{12}}R$.

\subsection{Dirac fermion}

To probe the fermionic response of the boundary theory via holography, we proceed to solve the Dirac fermion equation in the bulk curved spacetime of the charged Schr\"odinger black hole.
The action is 
\begin{equation} \label{Dirac}
S_{Dirac}=\int d^5x \sqrt{-g_{Ein}} i \bar{\psi} (e_{\hat{a}}^{\ \mu} \Gamma^{\hat{a}} \mathcal{D}_\mu-m) \psi,
\end{equation}

and its equation of motion(EOM) is
$
(e_{\hat{a}}^{\ \mu} \Gamma^{\hat{a}} \mathcal{D}_\mu-m) \psi=0,
$
with covariant derivative
\be
\mathcal{D}_\mu = \partial_\mu + \frac{1}{8} \eta_{\hat{a}\hat{c}} \omega^{\hat {c}}_{\hat{b}\mu}[\Gamma^{\hat{a}},\Gamma^{\hat{b}}] - i q A_\mu,
\ee
where gamma matrix of flat tangent space
$
\{\Gamma^{\hat{a}}, \Gamma^{\hat{b}}\} = 2\eta^{\hat{a}\hat{b}},
$
vielbeins $e_{\hat{a}}^{\ \mu}$ relates flat tangent space to curved spacetime,
$
g_{\mu\nu} e_{\hat{a}}^{\ \mu} e_{\hat{b}}^{\ \nu} = \eta_{\hat{a}\hat{b}}.
$
The spin connection is 
$
\omega^{\hat{c}}_{\hat{b}\mu} = 
e^{\hat{c}}_{ \nu}\partial_\mu e_{\hat{b}}^{\ \nu} +
\Gamma^\nu_{\ \sigma\mu} e^{\hat{c}}_{ \nu} e_{\hat{b}}^{\ \sigma}
$,
$\Gamma^\nu_{\ \sigma\mu}$ are the Christoffel symbols.
We choose the specific vielbein\footnote{For convenience, we rescale the coordinates to set $R=r_0=1$ from now on.},
\bea
e_{\hat{t}}^{\ t}=K^{\frac{1}{6}} \frac{\beta^2(f-1)}{2r^2f B}, \;\; e_{\hat{t}}^{\ \xi}= K^{\frac{1}{6}}\frac{f+1+2\sqrt{f/K}}{4r^2f B}, \;\;  e_{\hat{\xi}}^{\ t}= K^{\frac{1}{6}} B, \;\; e_{\hat{\xi}}^{\ \xi}= K^{\frac{1}{6}} \frac{B(f+1-2\sqrt{f/K})}{2\beta^2(f-1)}, \\
e_{\hat{x_1}}^{\ x_1}=K^{\frac{1}{6}} 1/r, \;\; e_{\hat{x_2}}^{\ x_2}=K^{\frac{1}{6}} 1/r,\;\; e_{\hat{r}}^{\ r}=K^{\frac{1}{6}} r\sqrt{f}.
\eea
with other unwritten components of $e_{\hat{a}}^{\ \mu}$ are zeros.
To simplify the Dirac equation calculation, here $B$ is a function of $r$ chosen ensuring the coefficient of $\Gamma^{\hat{t}}\Gamma^{\hat{\xi}}\Gamma^{\hat{r}}$ in the EOM is zero.
The boundary behavior of $B(r \rightarrow \infty)$ is a constant $B_b$ times $1/r$. We choose $B_b$ to be $1$.
Asymptotically,
$B(r \rightarrow \infty) \simeq \frac{1}{r} + \frac{3 \beta^2}{2r^3} + \frac{3(4-4\beta^2-3\beta^4)}{16 r^5} + \frac{-16+60 \beta ^2+24 \beta ^4+27 \beta ^6}{32 r^7}+ \frac{9 \left(48-64 \beta ^2-24 \beta ^4-48 \beta ^6-45 \beta ^8\right)}{256 r^9} + \cdots
$.
The near-horizon behavior of $B$ is $B_h/(r-1)$, where $B_h$ is a constant. Given $B_b=1$, we can numerically solve this equation to find $B_h$. 
In $5$D spacetime, each of $\Gamma^{\hat{a}}$ matrices has $4\times 4$ components, we choose to express them as follows:
\[
\Gamma_{\tau} = \left(
\begin{matrix}
0 & i\sigma_3 \\
i\sigma_3 & 0
\end{matrix}
\right), \quad
\Gamma_{y} =\left(
\begin{matrix}
0 & -iI \\
iI & 0
\end{matrix}
\right)
\]
\[
\Gamma_{x_1} = \left(
\begin{matrix}
0 & \sigma_2 \\
\sigma_2 & 0
\end{matrix}
\right),\quad
\Gamma_{x_2} = \left(
\begin{matrix}
0 & \sigma_1 \\
\sigma_1 & 0
\end{matrix}
\right),\quad
\Gamma_r = \left(
\begin{matrix}
I & 0 \\
0 & -I
\end{matrix}
\right)
\]
$\sigma_i$ are Pauli sigma matrices, $I$ is identity matrix. $\Gamma_t = \beta(\Gamma_\tau + \Gamma_y)$, $\Gamma_\xi = (-\Gamma_\tau + \Gamma_y)/(2\beta)$.
Rewrite the 4-component Dirac spinor field $\psi$ as:
\begin{equation}
\label{eq:ansatz}
\psi =\left(
\begin{matrix}
\psi_+\\
\psi_-
\end{matrix}
\right)
e^{-i\omega t +i\ell \xi + ik_1x_1 + ik_2x_2}
= (-gg^{rr})^{-1/4}\left(
\begin{matrix}
\phi_+\\
\phi_-
\end{matrix}
\right)e^{-i\omega t +i\ell \xi + ik_1x_1 + ik_2x_2}.
\end{equation}
where $\phi_+$ and $\phi_-$ are two-component spinors. This $(-gg^{rr})^{-1/4}$ factor eliminates a $\Gamma_{r}$ term in the Dirac equation, which is simplified to
\begin{equation} \label{eq-dirac}
\left(r\sqrt{f}\partial_r \mp m K^{-1/6}\right)\phi_\pm \pm 
\left(\pm u + v\sigma_3 + i\frac{k_1}{r} \sigma_2 + i\frac{k_2}{r} \sigma_1\right)
\phi_\mp= 0,
\end{equation}

where $u$ and $v$ are linear combinations of the
vielbein components
$e_{\hat{t}}^{\ t}$,$e_{\hat{t}}^{\ \xi}$,
$e_{\hat{\xi}}^{\ t}$ and $e_{\hat{\xi}}^{\ \xi}$:
\bea
u = 
K^{-\frac{1}{6}}\Bigg( \left(\omega+qA_t\right)\left( -\beta e_{\hat{t}}^{\;t} - \frac{1}{2\beta} e_{\hat{\xi}}^{\;t}  \right) + 
\left(\ell-qA_\xi\right)\left(\beta e_{\hat{t}}^{\;\xi} + \frac{1}{2\beta} e_{\hat{\xi}}^{\;\xi}  \right)\Bigg) 
\\
v =
K^{-\frac{1}{6}}\Bigg(\left(\omega+qA_t\right)\left( \beta e_{\hat{t}}^{\;t} - \frac{1}{2\beta} e_{\hat{\xi}}^{\;t}  \right) + 
\left(\ell-qA_\xi\right)\left(-\beta e_{\hat{t}}^{\;\xi} + \frac{1}{2\beta} e_{\hat{\xi}}^{\;\xi}  \right) \Bigg)  
\eea

By rotational symmetry of the boundary theory, we will work on the case $k_1 = 0$ and set $k_2 = k$ from here on.
We will write $\phi=(\phi_+,\phi_-)^{\text{T}}$, also its  
$\phi_+ = (y_+,\; z_+)^{\text{T}}$ 
and 
$\phi_- = (y_-,\; z_-)^{\text{T}}$ 
in the component form.
%

\section{Green's Function from Holography}  \label{sec3}
\subsection{Holographic dictionary} \label{dict}

We study the fermionic response of the boundary theory, by probing the Dirac fermion field in the bulk spacetime of Schr\"odinger black hole.
The holographic dictionary of source-response relation can be set up by reading the boundary action of Eq.(\ref{Dirac}). 
From \cite{Iqbal:2009fd,Henneaux:1998ch}, 
the variation of bulk action induces a boundary term $S_\partial=\int_{\partial \mathcal{M}} d^{3}x d\xi \sqrt{-g g^{rr}} \bar{\psi} \psi$. 
Therefore the relation between bulk field and its conjugate momentum are,
\be
\Pi_+=-\sqrt{-g g^{rr}} \bar{\psi}_-, \;\;\; \Pi_-=\sqrt{-g g^{rr}} \bar{\psi}_+ 
\ee
We can identify the source($\chi$) and response($\mathcal{O}$) from boundary(or UV) behavior of bulk field($\psi_{\pm}$) and momentum($\Pi_\pm$), from the holographic dictionary,
\be \label{gravity-gauge}
\exp[-S_{grav}{[\psi,\bar{\psi}](r \rightarrow \infty)}]=\langle \exp[\int d^{d+1}x (\bar{\chi} \mathcal{O} +\bar{\mathcal{O}} \chi )]\rangle_{QFT}
\ee

The source $\chi$ and bulk field $\psi$ are related by,
\be
\chi=\lim_{r \rightarrow \infty} r^{\frac{d+1}{2}-\nu_{\pm}} \psi
\ee
the response $\mathcal{O}$ and momentum $\Pi_\pm$ are related by,
\be
\mathcal {O}=-\lim_{r \rightarrow \infty} r^{\nu_{\pm}-\frac{d+1}{2}} \bar{\Pi}
\ee
where $\nu_\pm = \sqrt{(m\pm \frac{1}{2})^2+(\ell-q M_{o})^2}$ generically\footnote{specifically equal to $\sqrt{(m \pm 1/2)^2+(\ell+qQ\beta)^2}$ in our charged Schr\"odinger black hole case}, 
analogue to the result of \cite{Iqbal:2009fd}.
Here we show only the standard quantization(corresponding to source $A$ and response $D$ of \cite{Iqbal:2009fd}). 
The alternate quantization(corresponding to source $C$ and response $B$ of  \cite{Iqbal:2009fd}) can be done in the same manner\footnote{
For the alternate quantization,
\be
\chi=\lim_{r \rightarrow \infty} r^{\frac{d+3}{2}-\nu_{\pm}} \psi
\ee
the response $\mathcal{O}$ and momentum $\Pi_\pm$ are related by,
\be
\mathcal {O}=\lim_{r \rightarrow \infty} r^{\nu_{\pm}-\frac{d+3}{2}} \Pi
\ee
}. The Green's function $G_R$ is related to the ratio of $\mathcal{O}$ and $\chi$. \\

We now study the Dirac equation (\ref{eq-dirac}) in boundary UV asymptotic limit to extract $\mathcal {O}$ and $\chi$ from the coefficients of $\psi$ and $\Pi$, or equivalently related to $\phi_+$ and $\phi_-$ at $r \rightarrow \infty$. In this limit, (\ref{eq-dirac}) becomes
\[
\left(r\sqrt{f}\partial_r \mp mK^{-1/6} \right)\phi_\pm + 
\left(\frac{(\ell+qQ\beta)r}{2\beta}P_\pm + \frac{\mathcal{C}}{r}P_\pm + \frac{2\beta(\ell+qQ\beta)}{r}P_\mp \pm 
\frac{ik_1}{r} \sigma_2 \pm \frac{ik_2}{r}\sigma_1 
\right)\phi_\mp = 0
\]

where \[
\mathcal{C} = \frac{1}{8\beta^2}
\left(
-4qQ(1+\beta^2) +5(\ell+qQ \beta)\beta ^3-(\ell+q Q \beta )Q^2 \beta ^3 -8\beta\omega
\right), \;\;\; P_\pm = \frac{1 \pm \sigma_3}{2}
\]

\begin{multline} \label{UV1}
\phi_+ = 
\mathbf{S_1} \, r^{\nu_+-\frac{1}{2}}(A_1 + A_2 r^{-2}) + 
\mathbf{R_1} \, r^{-\nu_+-\frac{1}{2}}(\alpha_1 + \alpha_2 r^{-2}) + 
\mathbf{S_2} \, r^{\nu_-+\frac{1}{2}}(B_1 + B_2 r^{-2}) + 
\mathbf{R_2}\, r^{-\nu_-+\frac{1}{2}}(\beta_1 + \beta_2 r^{-2})+\dots
\end{multline}
\begin{multline} \label{UV2}
\phi_- = 
\mathbf{S_1}  \, r^{\nu_++\frac{1}{2}}(C_1 + C_2 r^{-2}) + 
\mathbf{R_1} \, r^{-\nu_++\frac{1}{2}}(\gamma_1 + \gamma_2 r^{-2}) + 
\mathbf{S_2} \, r^{\nu_--\frac{1}{2}}(D_1 + D_2 r^{-2}) + 
\mathbf{R_2} \, r^{-\nu_--\frac{1}{2}}(\delta_1 + \delta_2 r^{-2})+\dots
\end{multline}
here 
\be
\nu_\pm = \sqrt{(m \pm 1/2)^2+(\ell+qQ\beta)^2}.
\ee
Each of $\mathbf{S_1},\mathbf{S_2},\mathbf{R_1}, \mathbf{R_2}$ is a r-independent one-component multiplier, as the coefficient of the spinor\footnote{When doing numerics for this field redefinition, it is important to keep subleading term $C_2$ in the $\mathbf{S_1} \, r^{\nu_+ +\frac{1}{2}}(\dots+C_2 r^{-2}+\dots)$ series, since this $C_2$ term dominates $\mathbf{R_1} \, r^{-\nu_+-\frac{1}{2}}\alpha_1$ when $\nu_+ >\frac{1}{2}$, which is indeed our case in the numerical study. Thus here we keep the expansion for all four sets of solutions to the subleading orders.}.
There is a projection relation between the two-component spinors\footnote{Each of $A_1,A_2,C_1,C_2,\alpha_1, \alpha_2,\gamma_1, \gamma_2,B_1,B_2,D_1,D_2,\beta_1, \beta_2,\delta_1,\delta_2$ is a two-component spinor. List above totally there are sixteen two-component spinors. The spinors $C_1,C_2,\gamma_1$ and $\gamma_2$ are in the null space of $P_+$, the spinors $B_1,B_2,\beta_1$ and $\beta_2$ are in the null space of $P_-$. There are four independent sets of bases, each basis as a solution of Dirac EOM: the first set contains $A_1,A_2,C_1,C_2$ and its subleading terms, the second set contains $\alpha_1, \alpha_2,\gamma_1, \gamma_2$ and its subleadings, the third set contains $B_1,B_2,D_1,D_2$ and its subleadings, the fourth set contains $\beta_1, \beta_2,\delta_1,\delta_2$ and its subleadings.\label{Footnote6}}:
\bea 
\label{project1}
\mathbf{S_1}C_1=\frac{-(\ell+qQ\beta)}{2\beta(\nu_++m+\frac{1}{2})}P_- \mathbf{S_1}A_1, \;\;\; \mathbf{S_2}B_1=\frac{-(\ell+qQ\beta)}{2\beta(\nu_- -m+\frac{1}{2})}P_+ \mathbf{S_2}D_1,\\
\label{project2}
\mathbf{R_1}\gamma_1=\frac{-(\ell+qQ\beta)}{2\beta(-\nu_++m+\frac{1}{2})}P_- \mathbf{R_1}\alpha_1, \;\;\; \mathbf{R_2}\beta_1=\frac{-(\ell+qQ\beta)}{2\beta(-\nu_- -m+\frac{1}{2})}P_+\mathbf{R_2}\delta_1,
\eea

We now apply our holographic dictionary to identify the source and response from (\ref{UV1}), (\ref{UV2}). To read the boundary value, 
in the following we take $r \rightarrow \infty$ as UV limit. Consider the leading behavior of $\phi_-$ contribution, 
\be
\psi_-=(-g g^{rr})^{-1/4} \phi_-\simeq r^{-2} \phi_- \simeq \mathbf{S_1} C_1 r^{\nu_+ -3/2}
\ee
which corresponds to the source $\chi_-$,
\be
\chi_-=\lim_{r \rightarrow \infty} r^{\frac{3}{2}-\nu_{+}} \psi_-\simeq \mathbf{S_1}C_1
\ee
$\chi_-$ is proportional to $\mathbf{S_1}$.
The momentum field $\bar{\Pi}_-$ is
\be
\bar{\Pi}_-=\sqrt{-g g^{rr}} {\psi}_+ = {(-g g^{rr})}^{1/4} {\phi}_+  \simeq r^2 {\phi}_+ \simeq \mathbf{S_1} A_1 r^{\nu_+ +3/2} +\mathbf{R_1} \alpha_1 r^{-\nu_+ +3/2}+\dots
\ee
which corresponds to the response $\mathcal{O_-}$, 
\be
\mathcal{O_-}=\lim_{r \rightarrow \infty} r^{\nu_{+}-\frac{3}{2}} \bar{\Pi}_- \simeq \mathbf{R_1} \alpha_1 
\ee
We take the asymptotic constant term on the UV boundary.
$\mathcal{O_-}$ is proportional to $\mathbf{R_1}$.\\

On the other hand, we can go through the same logic again, though consider the leading behavior of $\phi_+$ contribution, 
\be
\psi_+=(-g g^{rr})^{-1/4} \phi_+\simeq r^{-2} \phi_+ \simeq \mathbf{S_2} B_1 r^{\nu_- -3/2}
\ee
which corresponds to the source $\chi_+$,
\be
\chi_+=\lim_{r \rightarrow \infty} r^{\frac{3}{2}-\nu_{-}} \psi_+\simeq \mathbf{S_2}B_1
\ee
$\chi_+$ is proportional to $\mathbf{S_2}$.
The momentum field $\Pi_+$ is
\be
\bar{\Pi}_+=-\sqrt{-g g^{rr}}  {\psi}_- =- {(-g g^{rr})}^{1/4}  {\phi}_-  \simeq - r^2 {\phi}_- \simeq - \mathbf{S_2} D_1 r^{\nu_- +3/2} -\mathbf{R_2} \delta_1 r^{-\nu_- +3/2}+\dots
\ee
which corresponds to the response $\mathcal{O_+}$, 
\be
\mathcal{O_+}=-\lim_{r \rightarrow \infty} r^{\nu_{-}-\frac{3}{2}} \bar{\Pi}_+ \simeq \mathbf{R_2} \delta_1 
\ee
$\mathcal{O_+}$ is proportional to $\mathbf{R_2}$. Now we derive $\mathbf{S_1},\mathbf{S_2}$ are identified as sources, $\mathbf{R_1},\mathbf{R_2}$ are identified as responses for this standard quantization. A similar argument works for the alternative quantization by taking the subleading terms of the bulk field and its conjugate momentum, we leave this detail in the Appendix \ref{append:alternate}.\\


In addition to our above holographic dictionary, we provide another intuitive argument on identifying source and response. We notice the boundary action $\bar{\psi} \psi$, due to the $\Gamma_{\tau}$ form, it couples the first component of the spinor $\psi$ to the third component of $\psi$, while couples the second component of $\psi$ to the fourth component of $\psi$. Both two-point function or $\bar{\psi} \psi$ shows $r^{2\nu_+}$ scaling in \cite{Akhavan:2009ns}\cite{Leigh:2009ck} and our work. 
However, the $r^{2\nu_-}$ scaling is only seen in our and \cite{Leigh:2009ck}'s Green's functions. 
We will perform a more constructive comparison with \cite{Akhavan:2009ns},\cite{Leigh:2009ck} and a pure Schr\"odinger Green's function computation via our dictionary in Appendix \ref{append:pureSchr}.
Lastly, we are aware that the detailed construction of Schr/NRCFT holographic dictionary involves nontrivial holographic renormalization\cite{Leigh:2009ck,Guica:2010sw,vanRees:2012cw}. 
Our work here only follows the strategy in Ref.\cite{Liu:2009dm,Iqbal:2009fd} constructing the source-response holographic dictionary. 
The rigorous holographic 
renormalization for spinors is the future step to justify the complete dictionary for the Green's function.

\subsection{Source and response from UV expansion}

To extract the data of source and response, 
we define a converting matrix ${\Cv}$ as a function of $r$(see Appendix \ref{append:Green} and a shared program code through a URL link), and a set of functions $\mathbf{S_1}(r), \mathbf{S_2}(r), \mathbf{R_1}(r), \mathbf{R_2}(r)$ satisfies
\be \label{phi-conv}
\phi(r)=\left(
\begin{matrix}
\phi_+(r)\;
\phi_-(r)\\
\end{matrix}
\right)^{\text{T}} 
 = 
\left(
\begin{matrix}
y_+(r)\;
z_+(r)\;
y_-(r)\;
z_-(r)
\end{matrix}
\right)^{\text{T}} 
 = {\Cv} \cdot
\left[
\begin{matrix}
\mathbf{S_1}(r)\;
\mathbf{R_1}(r)\;
\mathbf{S_2}(r)\;
\mathbf{R_2}(r)
\end{matrix}
\right]^{\text{T}} 
\ee

By this field definition, neatly $\mathbf{S_1}(r), \mathbf{S_2}(r), \mathbf{R_1}(r), \mathbf{R_2}(r)$ approach to $\mathbf{S_1}, \mathbf{S_2}, \mathbf{R_1}, \mathbf{R_2}$ as $r \rightarrow \infty$.
Due to projection, we find the spinors have the properties $c_{1+} = c_{2+} = \gamma_{1+} =\gamma_{2+} = b_{1-} = b_{2-} = \beta_{1-} = \beta_{2-} = 0$. 
To deal with the standard quantization, from the lesson of Sec.\ref{dict}, we choose $c_{1-}=1$ to compute the first set of bases, $\alpha_{1-}=1$ to compute the second set of bases, $b_{1+}=1$ to compute the third set of bases,  $\delta_{1+}=1$ to compute the fourth set of bases. 
Each of four independent bases in (\ref{UV1}) and (\ref{UV2})(equivalently in \ref{cov}, see Footnote.\ref{Footnote6}), can be determined by a free parameter, thus totally four free parameters. 
Now the four free parameters for four independent bases are $\mathbf{S_1}, \mathbf{S_2}, \mathbf{R_1}, \mathbf{R_2}$. Argue from Sec.\ref{dict}, for the standard quantization,
the source terms are $\mathbf{S_1}C_1,\mathbf{S_2}B_1$, with their corresponding response terms $\mathbf{R_1}\alpha_1,\mathbf{R_2}\delta_1$ respectively. Our choice of spinor $C_1^T=(0,1)$ and its coupled spinor $\alpha_1^T=(\alpha_{1+}, 1)$ justifies that coefficient $\mathbf{S_1}$ is exactly a source and $\mathbf{R_1}$ is its response. Our choice of spinor $B_1^T=(1,0)$ and its coupled spinor $\delta_1^T=(1,\delta_{1-})$ justifies that coefficient $\mathbf{S_2}$ is exactly a source and $\mathbf{R_2}$ is its response
\footnote{For the alternate quantization, we should alternatively choose $a_{1-}=1$ to compute the first set of bases, $\gamma_{1-}=1$ to compute the second set of bases, $d_{1+}=1$ to compute the third set of bases, $\beta_{1+}=1$ to compute the fourth set of bases.
From Sec.\ref{dict}, the source terms are $\mathbf{S_2}D_1,\mathbf{S_1}A_1$; with their corresponding response terms $\mathbf{R_2}\beta_1,\mathbf{R_1} \gamma_1$ respectively.  
Our choice of spinor $A_1^T=(a_{1+},1)$ and its coupled spinor $\gamma_1^T=(0, 1)$ justifies that coefficient $\mathbf{S_1}$ is exactly a source and $\mathbf{R_1}$ is its response. Our choice of spinor $D_1^T=(1,d_{1-})$ and its coupled spinor $\beta_1^T=(1,0)$ justifies that coefficient $\mathbf{S_2}$ is exactly a source and $\mathbf{R_2}$ is its response.}.\\

\subsection{IR behavior and the In-falling boundary condition} \label{sec3-IR}

To determine the near horizon initial condition of Dirac equation, here we deal with IR behavior and solve the in-falling boundary condition at zero temperature, $Q=\sqrt{2}$. 
Consider the small $\epsilon$ expansion of the equations, where $r=1+\epsilon$. 
The equation for $B(r)$ becomes $\frac{B'}{B} = -1/\epsilon$.
Thus, we take the behavior of $B(r)$ near horizon as $B_h/\epsilon$, where $B_h$ is another constant.
$\lim_{r\to 1} (r-1)B(r) = B_h$
For the later use, we define 
\be
\tilde{\omega} = \frac{\ell}{2\beta} + \beta\omega   
\ee
$\tilde{\omega}$ 
is the coefficient of $\tau$ in the exponent dependence of Eq.(\ref{eq:ansatz}). 
We find that the behavior of $f$, $u$,$v$ in the IR is given by 
\be
f \to 12\epsilon^2,\;\; u\to\frac{i \tilde{\omega} }{\epsilon} U,\;\; v\to \frac{i \tilde{\omega}}{\epsilon} V 
\ee
with $U \equiv \frac{i}{4\sqrt{3}}\left( \frac{2\sqrt{3} B_h}{\beta^2} -  \frac{\beta^2}{2\sqrt{3} B_h} \right)$ and $V \equiv \frac{i}{4 \sqrt{3}}\left( \frac{2\sqrt{3} B_h}{\beta^2} + \frac{\beta^2}{2\sqrt{3} B_h} \right)$.
Dirac equation near horizon and its infalling wave function ansatz are, 
\bea
\epsilon^2\phi_\pm' &=& \frac{-i}{2\sqrt{3}}\tilde{\omega}(U\pm V\sigma_3)\phi_\mp \label{eq:IReom} \\ 
\phi_\pm &\propto& e^{+i\tilde{\omega}/(12\epsilon)}  \label{eq:infalling} 
\eea
The exponent of wave function $\phi_\pm$ is chosen to be $+$ sign, in order to combined with Eq.(\ref{eq:ansatz}) to be $e^{-i \tilde{\omega} \tau +i\tilde{\omega}/(12\epsilon)}$ infalling into the black hole\footnote{
The $12$ factor appears here origins from the near horizon geometry $\AdS_2$.}.
The infalling condition is obtained by plugging Eq.(\ref{eq:infalling}) into Eq.(\ref{eq:IReom}), where the subscript $H$ stands for values at the horizon. 
\begin{equation}
\phi_+\vert_H = (U+V\sigma_3)\phi_-\vert_H
\label{bc}
\end{equation}
The infalling condition for spinors has two linear independent choices, the first set is $\phi_{-,1}=(1, 0)$ thus $\phi_{+,1}=(U+V, 0)$, and the second set is $\phi_{-,2}=(0,1)$ thus $\phi_{+,2}=(0,U-V)$. 
Therefore this gives two independent sets of initial conditions at horizon for $\mathbf{S_1}(r), \mathbf{R_1}(r), \mathbf{S_2}(r), \mathbf{R_2}(r)$, which we introduce one more upperindices $1,2$ to distinguish the first and the second sets:
\bea
\left[
\mathbf{S}_{\mathbf{1}}^1(r)\;
\mathbf{R}_{\mathbf{1}}^1(r)\;
\mathbf{S}_{\mathbf{2}}^1(r)\;
\mathbf{R}_{\mathbf{2}}^1(r)
\right]^{\text{T}}\vert_H &=
{\Cv}^{-1} \vert_H \cdot \begin{pmatrix} \phi_{+,1}& \phi_{-,1} \end{pmatrix}^{\text{T}} 
 \vert_H
&={\Cv}^{-1}\vert_H\cdot 
\begin{pmatrix}
U+V & 0 &1 &0
 \end{pmatrix}
^{\text{T}} \nonumber\\
\left[
\mathbf{S}_{\mathbf{1}}^2(r)\;
\mathbf{R}_{\mathbf{1}}^2(r)\;
\mathbf{S}_{\mathbf{2}}^2(r)\;
\mathbf{R}_{\mathbf{2}}^2(r)
\right]^{\text{T}}\vert_H &=
{\Cv}^{-1} \vert_H\cdot \begin{pmatrix} \phi_{+,2}&\phi_{-,2} \end{pmatrix} ^{\text{T}}\vert_H
&= {\Cv}^{-1}\vert_H\cdot
\begin{pmatrix}
0 & U-V & 0 & 1
\end{pmatrix}
^{\text{T}} \nonumber
\eea

More conveniently in matrix form,
\begin{equation*}
\mathbf{S}(r) = \left[
\begin{matrix}
\mathbf{S}_{\mathbf{1}}^1(r) & \mathbf{S}_{\mathbf{1}}^2(r) \\
\mathbf{S}_{\mathbf{2}}^1(r) & \mathbf{S}_{\mathbf{2}}^2(r)
\end{matrix}
\right], \quad
\mathbf{R}(r) = \left[
\begin{matrix}
\mathbf{R}_{\mathbf{1}}^1(r) & \mathbf{R}_{\mathbf{1}}^2(r) \\
\mathbf{R}_{\mathbf{2}}^1(r) & \mathbf{R}_{\mathbf{2}}^2(r)
\end{matrix}
\right],
\end{equation*}

\subsection{Green's function}

Green's function of the boundary theory is defined to be the ratio between source matrix $\mathbf{S}(r)$ and response matrix $\mathbf{R}(r)$. 
Thus we define $\mathbf{G}(r)$ based on $\mathbf{R}(r) = \mathbf{G}(r) \mathbf{S}(r)$, 
and evaluate $\mathbf{G}(r)$ at $r\rightarrow \infty$,
to read the $2\times 2$ matrix Green's function $\mathbf{G}$ of the boundary theory\footnote{There is no extra $\Gamma^\tau$ factor multiplied with $G(r)$ for this Green's function, because in our dictionary sources and responses are related to the coefficients of two-component spinors, instead of spinors itself.},
\be \label{Ginfty}
\mathbf{G}=\lim_{r\rightarrow \infty} \mathbf{G}(r)=\lim_{r\rightarrow \infty} \mathbf{R}(r) \mathbf{S}(r)^{-1}
\ee
We derive the EOM of $\mathbf{G}(r)$ in the bulk gravity(see Appendix \ref{append:Green}),
and solve this EOM with the initial condition:$
\label{eq-initialG}
\quad \mathbf{G}(r) \vert_H = \mathbf{R}(r)\vert_H \cdot \mathbf{S}(r)^{-1}\vert_H
$.
to obtain physical results of Eq.(\ref{Ginfty}).

\section{Spectral Function $A(k,\omega)$} \label{sec4}

We now equip with the holography tool developed in Sec.\ref{sec2},\ref{sec3}.  
The original questions driving our interests are: what is the nearly ground state of this fermionic system under asymptotic NRCFT background at zero-temperature? Will there be a Fermi surface? Will Fermi surface collapse, destabilized by tuning non-temperature parameters (such as background density $\beta$)? Will this realize certain quantum phase transition of fermionic liquids?
With the holographic dictionary for Green's function, we proceed to study these questions.
We focus on $Q=\sqrt{2}$ as zero temperature phase.\\
  
The spectral function $A(k,\omega)$ of this boundary system can be determined by the imaginary part of Green's function.
In $2+1$ D boundary theory with 
$2 \times 2$ matrix $\mathbf{G}$, we should take eigenvalues of $\mathbf{G}$, namely,
$\Img[\mathbf{G(r \rightarrow \infty)}_{eigenvalues}]$
\footnote{\label{footnote:akw} In principle, the spectral function is written as,
\be A(k,\omega)=-\frac{1}{\pi} \Img[G].
\ee 
The usual ARPES(Angle Resolved Photo Emission Spectroscopy) data\cite{Damascelli:2003bi} sum rule is $ \int^{\infty}_{-\infty} A(k,\omega) d \omega=1$\cite{Damascelli:2003bi}. 
This ARPES sum rule holds in a non-relativistic system.
The relativistic version of sum rule is written as, see Ref.[\onlinecite{Kapusta:2006pm}],
$ \int^{\infty}_{-\infty} \omega A(k,\omega) d \omega=1$.
In the context of gauge-gravity duality, the modification of ARPES sum rule has been studied\cite{Gulotta:2010cu},
eg. see a comment at Eq.(5.27) of Ref.[\onlinecite{Gulotta:2010cu}].
To produce any of the above sum rules, we comment a subtlety in Schr\"odinger holography. When we relate the $d+3$-D gravity theory to a $d+1$-D boundary theory via Eq.(\ref{gravity-gauge}), the constant factor $\int d\xi\equiv L_\xi$ needs to be absorbed into $\int d^{d+1}x (\bar{\chi} \mathcal{O} +\bar{\mathcal{O}} \chi)$, this gives an extra constant factor for source field or response field. Namely, the $A(k,\omega)$ may be different from $\Img[\mathbf{G}]$ with another extra constant factor. 
This factor should be important when justifying spectral density sum rule, $\int^{\infty}_{-\infty} A(k,\omega) d \omega=1$. 
The exact value of our $\Img[\mathbf{G}]$ is less informative, only the relative height of $\Img[\mathbf{G}]$ has physical indication.
In addition, we are aware that there is an alternative proposal to study the trace part\cite{Cubrovic:2009ye}, $\Img[\Tr[\mathbf{G(r \rightarrow \infty)}]]$.
}. 
%
%

\subsection{Fermi surface} \label{sec:FS}

This non-relativistic fermionic system has five parameters, conformal dimensions $\Delta_\pm(\nu_\pm)$, temeprature $T$, chemical potential(of background) $\mu_Q$, particle number eigenvalue or mass $M$, and background density $\beta$.
We first study the background density at $\beta=1/\sqrt{2}$ at $T=0$. 
The gauge-invariant mass operator $M \equiv \ell-q M_{o} =\ell+qQ\beta $ is fixed to be 1/10. The Dirac fermion charge $q=1$, its mass is chosen to be $m=1/10$, nonzero value in order to avoid scaling dimension $\nu_{\pm}$ degeneracy and extra logarithmic term in UV expansion. 

\begin{figure}[!h]
\centerline{ (a)\epsfig{figure=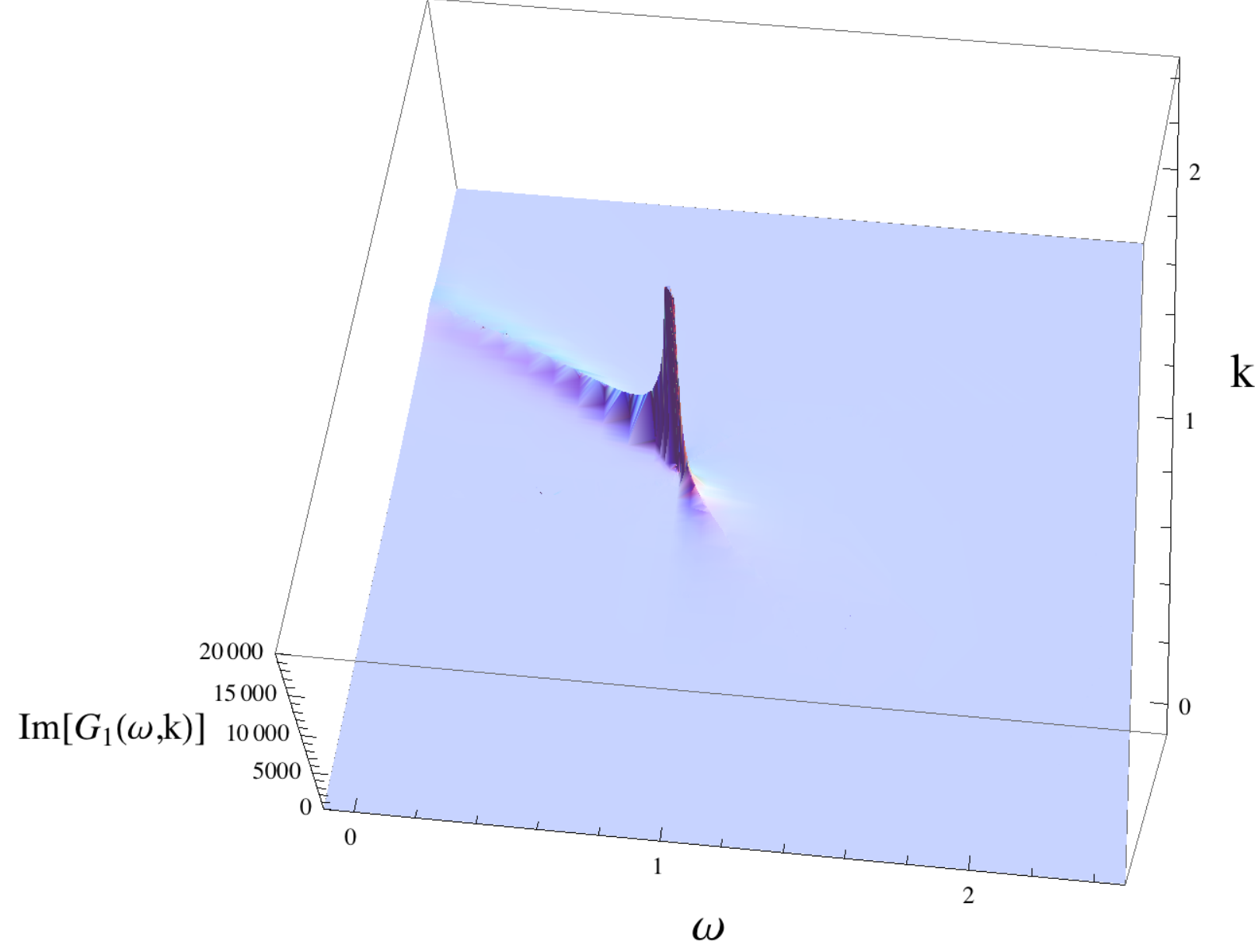, width=3.8 in}   (b)\epsfig{figure=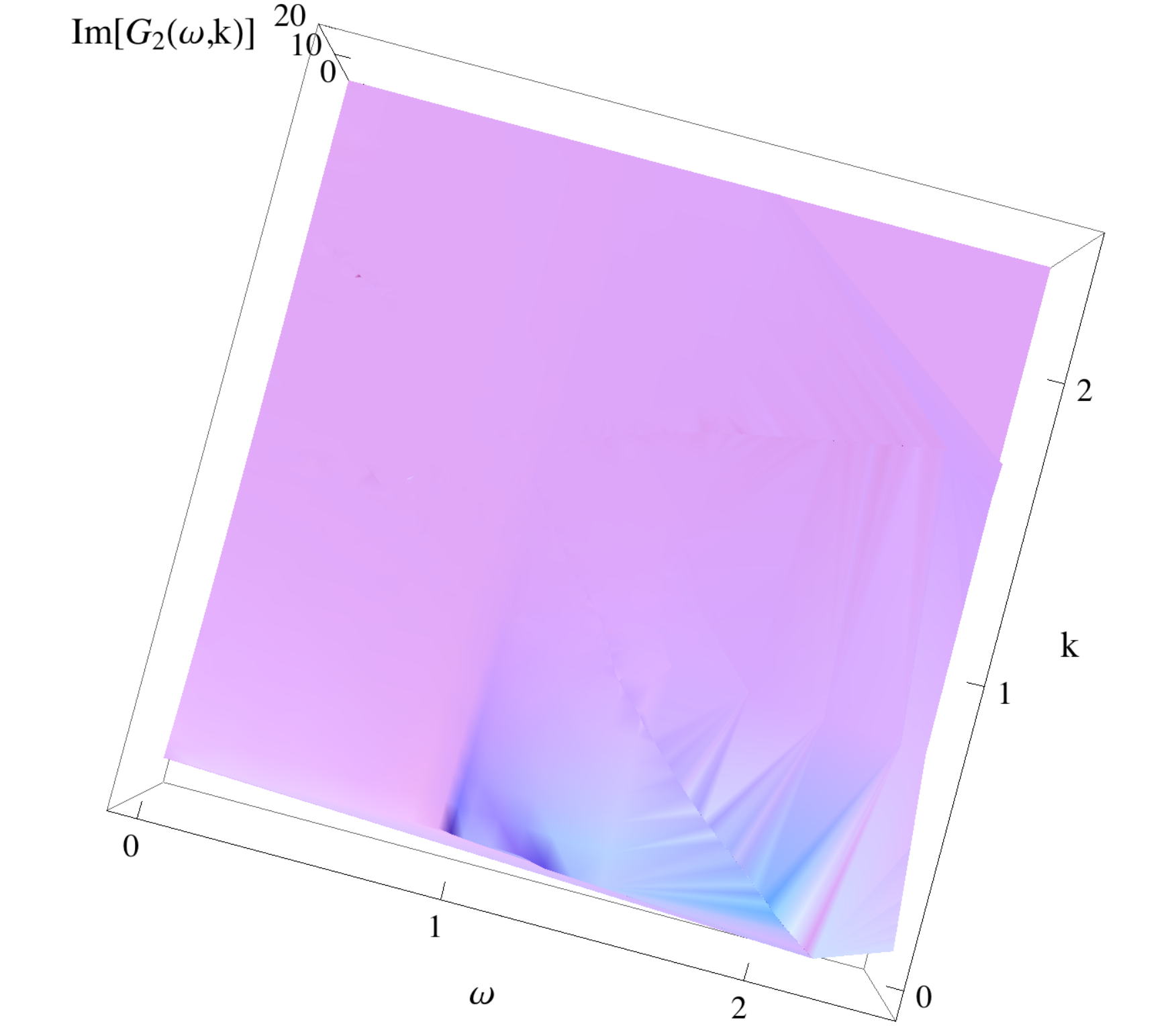, width=3.2 in}} 
\caption{At $\beta=1/\sqrt{2}$, (a) the imaginary part of Green's function, $\Img[G_1]$ as a function of $\omega$ and $k$. A sharp quasiparticle-like pole at $\omega_F=0.8984,k_F=1.3169$ indicates a well-defined Fermi surface. The pole indicates infinite lifetime stable quasiparticle at $k_F$. Notice the main branch of dispersion goes into $\omega<\omega_F$ and $k>k_F$, a hole-like excitation. While in \cite{Liu:2009dm},  their main branches of dispersion goes into $\omega>\omega_F$ and $k>k_F$, a particle-like excitation. 
(b) the imaginary part of Green's function, $\Img[G_2]$ as a function of $\omega$ and $k$, it is more or less featureless, except a wedge-like structure.} 
\label{fig:Gkw} 
\end{figure}

Similar to \cite{Liu:2009dm}, among two eigenvalues (say $\mathbf{G}_1, \mathbf{G}_2$) 
one of the eigenvalues, $\mathbf{G}_1$ with its imaginary $\Img[\mathbf{G}_1]$ has shown a pole-like structure (see \figref{fig:Gkw}(a) ), thus is picked for detailed studies in our analysis.
The other eigenvalue $\mathbf{G}_2$ with its imaginary $\Img[\mathbf{G}_2]$, only shows less-distinguished wedge-like structure  (see \figref{fig:Gkw}(b) ),
which appears to be less interesting physically. 
%
%
Following \cite{Liu:2009dm}, we focus on study one of these eigenvalues, $\mathbf{G}_1$.  
Below we will simply abbreviate $\Img[\mathbf{G}_1]$,$\Rea[\mathbf{G}_1]$ as $\Img G_1, \Rea G_1$.
We find there is a sharp pole on $\Img[G_1]$ at $\omega_F=0.8984,k_F=1.3169$, indicating a stable quasiparticle like excitation at Fermi-momentum $k_F$. This indicates a well-defined Fermi surface.
Normally the location of Fermi surface on $\omega$-axis is shifted by chemical potential $\mu$, one redefines $\bar{\omega}=\omega-\mu$ thus $\bar{\omega}=0$ has the Fermi surface. In our case, $\omega_F$ is shifted by the presence of $\xi$-momentum $\ell$, this can be realized from the fact that the location of Fermi surface is determined mainly from the low energy IR physics. which in the bulk gravity corresponds to near horizon region. Thus, instead of using boundary time coordinate $t$ and its coupled conjugate energy $\omega$, we identify the near-horizon time coordinate $\tau$ and its conjugate energy $\tilde{\omega}$. When $\tilde{\omega} =(\frac{\ell}{2\beta}+\beta \omega) \vert_{\omega_F}=0$, namely $\omega_F=-\ell/(2\beta^2)$, its value indicates the pole location of a Fermi surface.
Denote $k_\perp\equiv |k-k_F|$, we find near the quasiparticle like peak has scalings, 
 \bea
 \omega_*(k_\perp) &\sim& k_\perp^z,\;\;\;\;  z  \simeq  1.14\\
 \Img[G_1(\omega_*(k_\perp),k_\perp)] &\sim& k_\perp^{-\alpha},\;\;\;  \alpha \simeq 1.00
 \eea

\begin{figure}[!h]
\centerline{ (a)\epsfig{figure=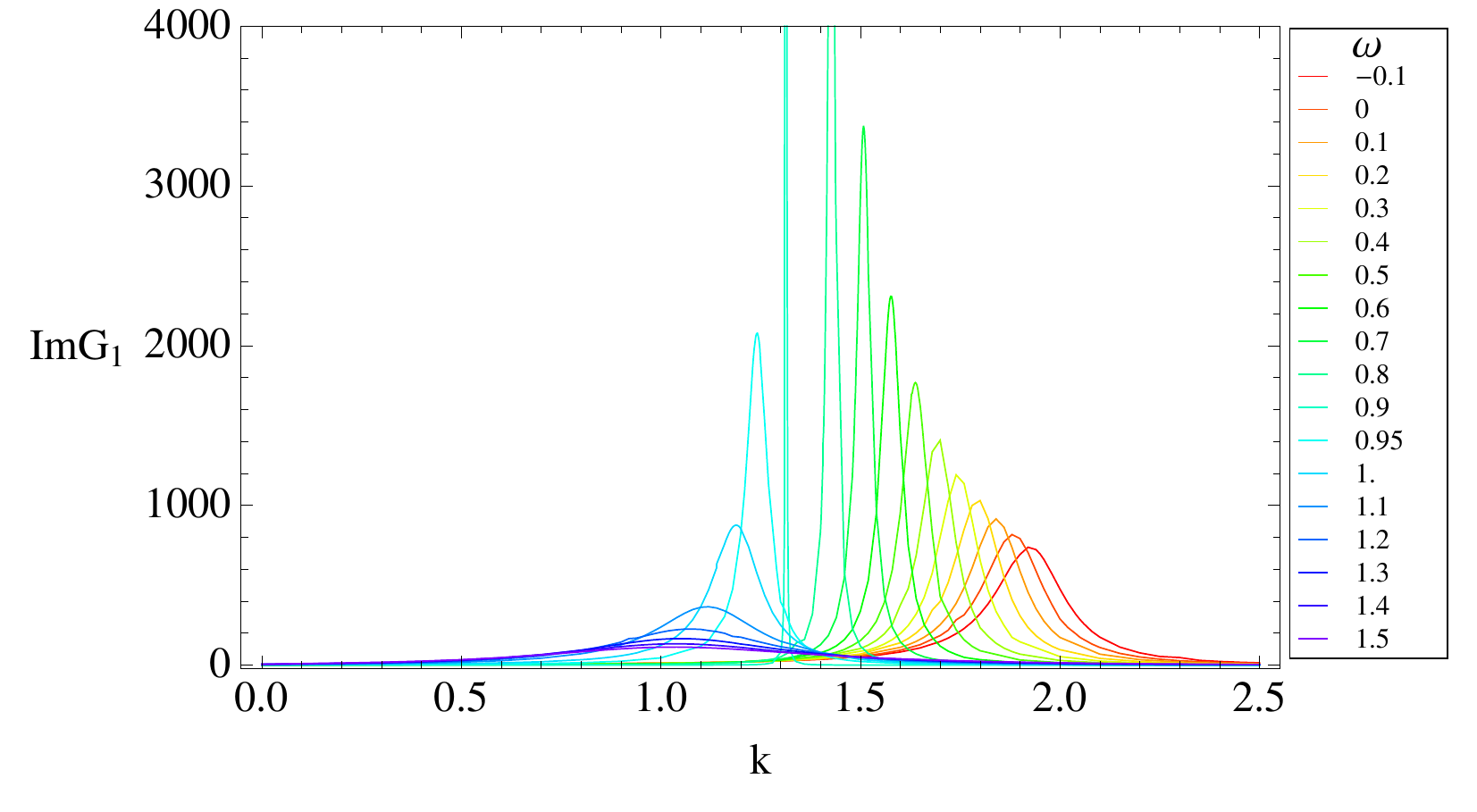, width=3.6 in} (b)\epsfig{figure=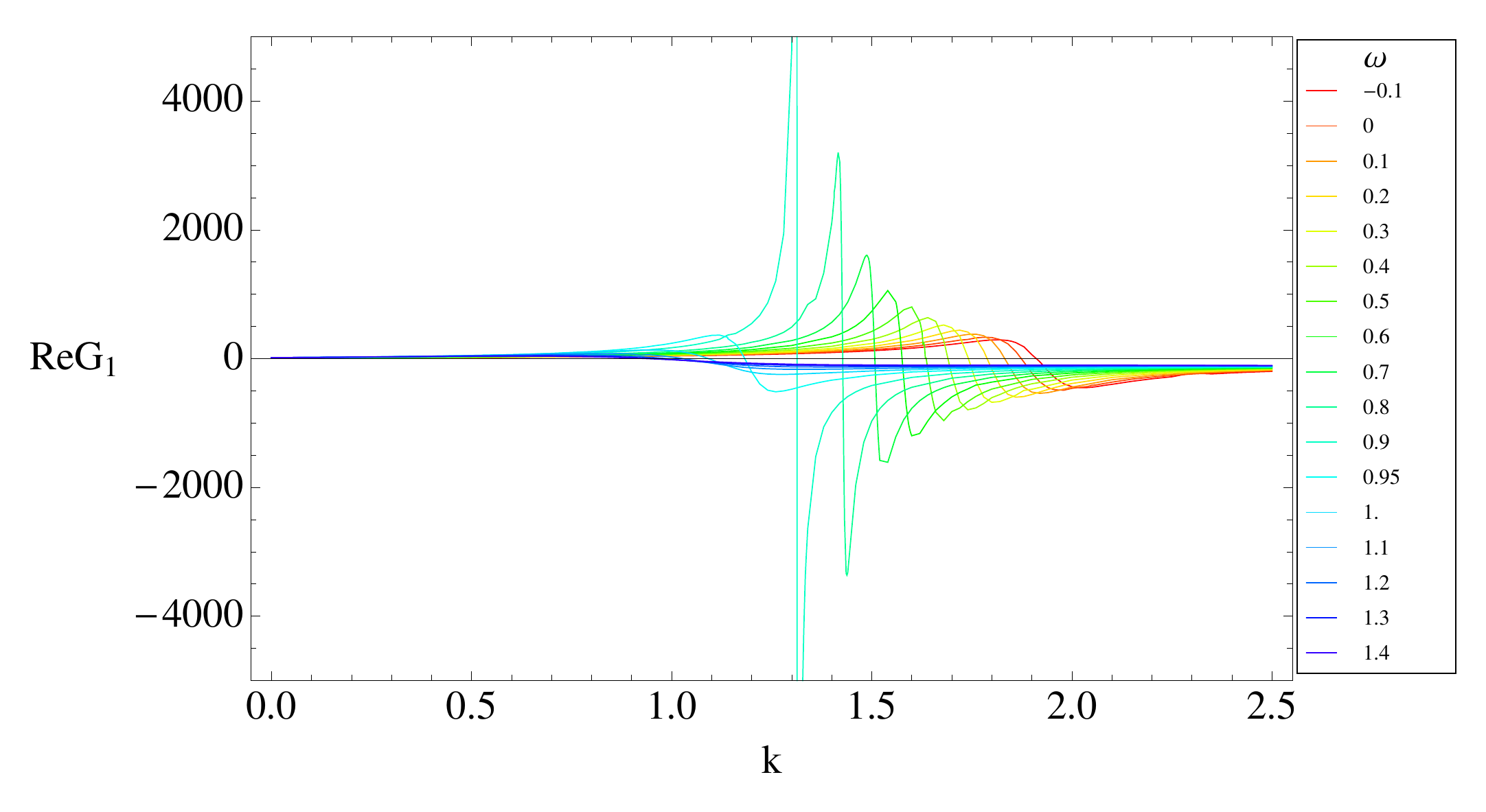, width=3.6 in} }
\caption{At $\beta=1\sqrt{2}$, (a)The imaginary part of Green's function, with a pole. (b)The real part of Green's function switches sign crossing zero value at a specific $\omega$ near the peak of $\Img[G_1]$.} 
\label{fig:img1beta_1_sqt2} 
\end{figure}

\begin{figure}[!h]
\centerline{\epsfig{figure=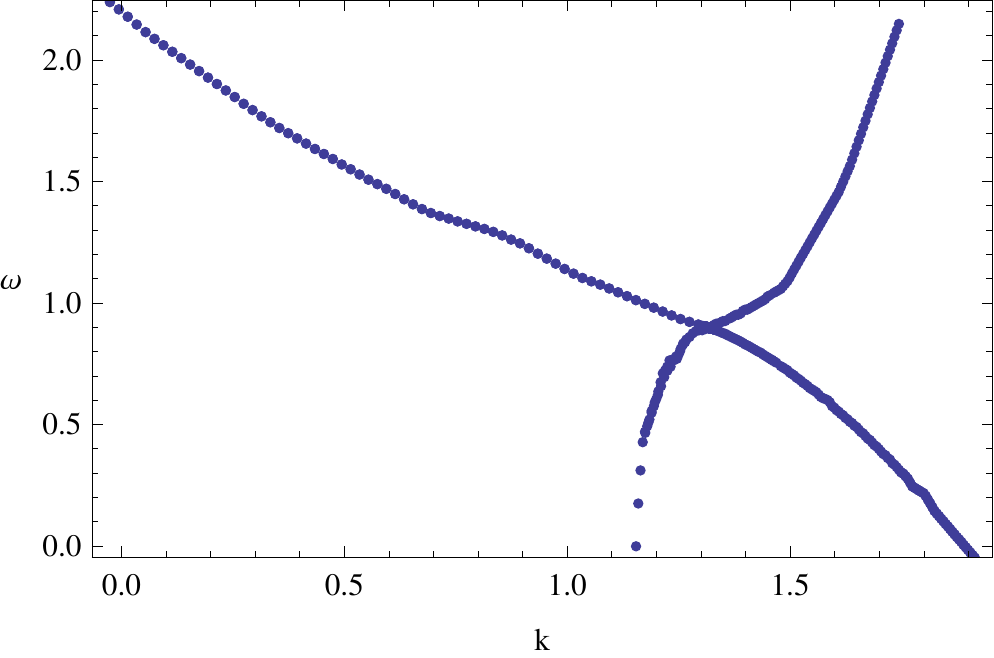, width=3.2 in}} 
\caption{The dispersion relation traces the pole into four branches on $(k,\omega)$ plane in a large scale. The asymmetric behavior indicates a particle-hole asymmetry.} 
\label{fig:Akw} 
\end{figure}

In \figref{fig:img1beta_1_sqt2}, we show the imaginary part $\Img[\mathbf{G}_1]$ and real part $\Rea[\mathbf{G}_1]$ of Green's function,
see the location of Fermi surface indicates a pole in $\Img[\mathbf{G}_1]$ and switches the sign of $\Rea[\mathbf{G}_1]$. 
The main branch of dispersion goes into $\omega<\omega_F$ and $k>k_F$, which is a hole-like excitation
The result is different from \cite{Liu:2009dm}, where their main branches goes into a particle-like excitation with $\omega>\omega_F$ and $k>k_F$.
In \figref{fig:Akw}, the dispersion relation shows particle-hole asymmetry in large scale, though close to $(\omega_F,k_F)$, it gives a unique dynamical exponent $z$.

\subsection{Comparison to Landau Fermi liquid theory and Senthil's scaling ansatz}

To better understand the physics of Green's function $G(k,\omega)$, we now study the functional form of $G(k,\omega)$ in terms of two different classes. Both classes hold under general arguments. The first class is Landau Fermi liquid theory, which holds for weak coupling system, where the free fixed point is still a good description of the system. The second class is even more general based on scaling ansatz for non-Fermi liquid theory and critical Fermi surface, proposed by Senthil\cite{Senthil,senthil:0804}.
In Landau Fermi liquid(LFL) theory, the retarded Green's function is of the form,
\be
G(k,\omega)=\frac{1}{\omega-\xi_k-\Sigma(\omega ,k)}=\frac{1}{\omega-(\xi_k+\Rea \Sigma(k,\omega))-i \Img\Sigma(k,\omega)}=\frac{Z}{(\omega-\omega_F)-\tilde{\xi_k}+\frac{i}{2\tilde{\tau_k}}}
\ee
$\Sigma(\omega ,k)$ is the particle irreducible retarded self-energy. $\xi_k\equiv \epsilon_k-\mu$, is the excitation around the original chemical potential. The condition $\xi_{k_F}+\Rea\Sigma(k_F,\omega_F)=\omega_F$ to define renormalized Fermi-momentum $k_F$. The final form is obtained by expanding $\xi_{k}+\Rea\Sigma(k,\omega)$ around $(k_F,\omega_F)$, 
with the definition of quasiparticle residue $Z$, with $Z^{-1} \equiv1-\frac{\partial}{\partial \omega} \Rea\Sigma(k=k_F,\omega=\omega_F)$, also $\tilde{\xi_k}\equiv(k-k_F)Z\frac{\partial}{\partial k}(\xi_{k_F}+\Rea\Sigma(k_F,\omega_F))\equiv v(k-k_F)$, and quasiparticle decay rate ${1}/{\tau_k}\equiv-2Z \Img \Sigma(k,\omega)$. 
The specific LFL form we use to fit our Green's function is 
\be \label{LFL}
G=\frac{Z}{(-(\omega-\omega_F)-v(k-k_F))-i \gamma(\omega)} 
\ee
with our quasiparticle self-energy ansatz as $\gamma(\omega)=\kappa(\omega-\omega_F)^n$, where $\kappa$ is some real constant, LFL has $n=2$. We will take general $n$ for fitting ansatz.
{We flip the sign of $(\omega-\omega_F)$ to have a hole-like dominant excitation as Fig.\ref{fig:Akw} suggests.}\\

The scaling ansatz proposed by Senthil\cite{Senthil,senthil:0804} based on general arguments, has the form at $T=0$,
\be
G=c_0 (k-k_F)^{-\alpha} F_0(\frac{c_1 (\omega-\omega_F)}{(k-k_F)^z})
\ee
for better fitting we will be forced to choose $c_0$ and $c_1$ their values on two sides $\omega>\omega_F$ and $\omega< \omega_F$ differently to reflect particle-hole asymmetry.\\

\begin{figure}[!h]
\centerline{ \epsfig{figure=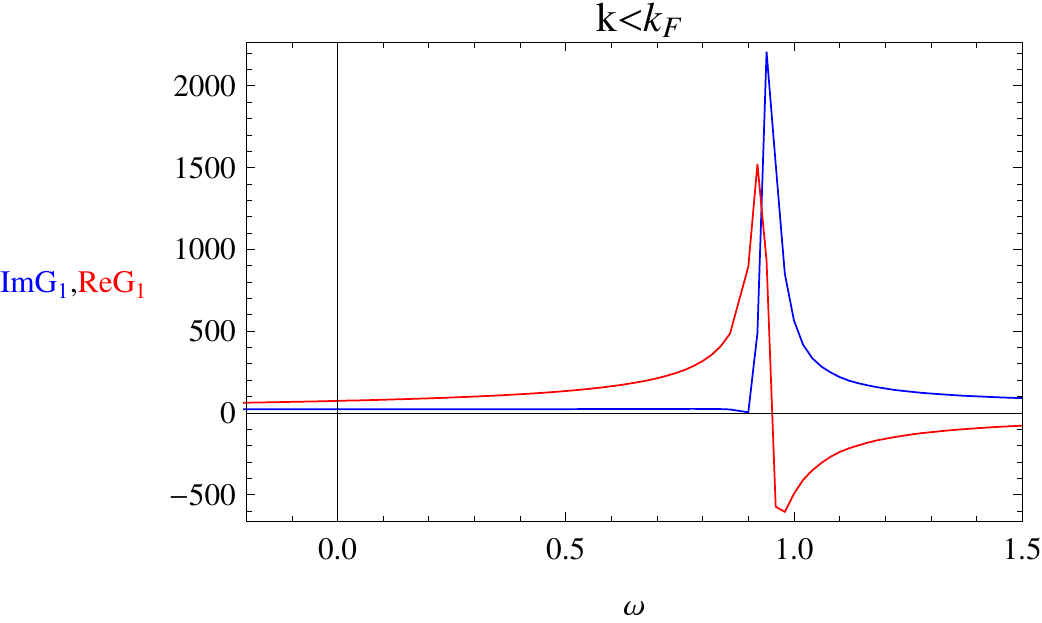, width=3.2 in}} 
\caption{$k=621/500<k_F$. Blue curve is for $\Img G_1$, red curve is for $\Rea G_1$.} 
\label{fig:data1} 
\end{figure}

\begin{figure}[!h]
\centerline{ (a)\epsfig{figure=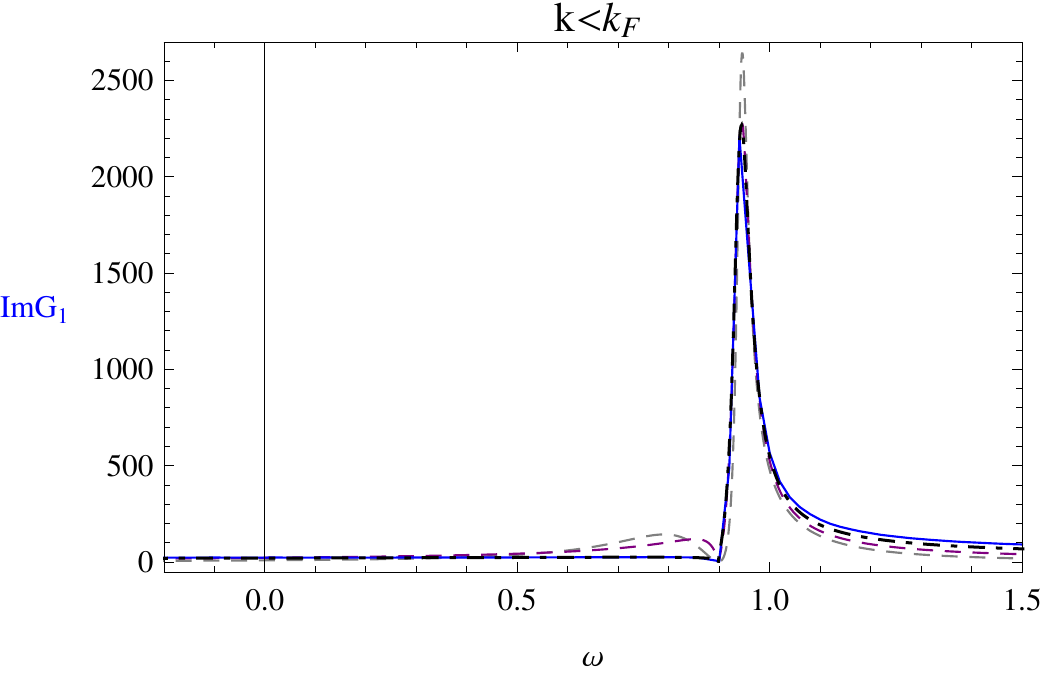, width=3.6 in} (b)\epsfig{figure=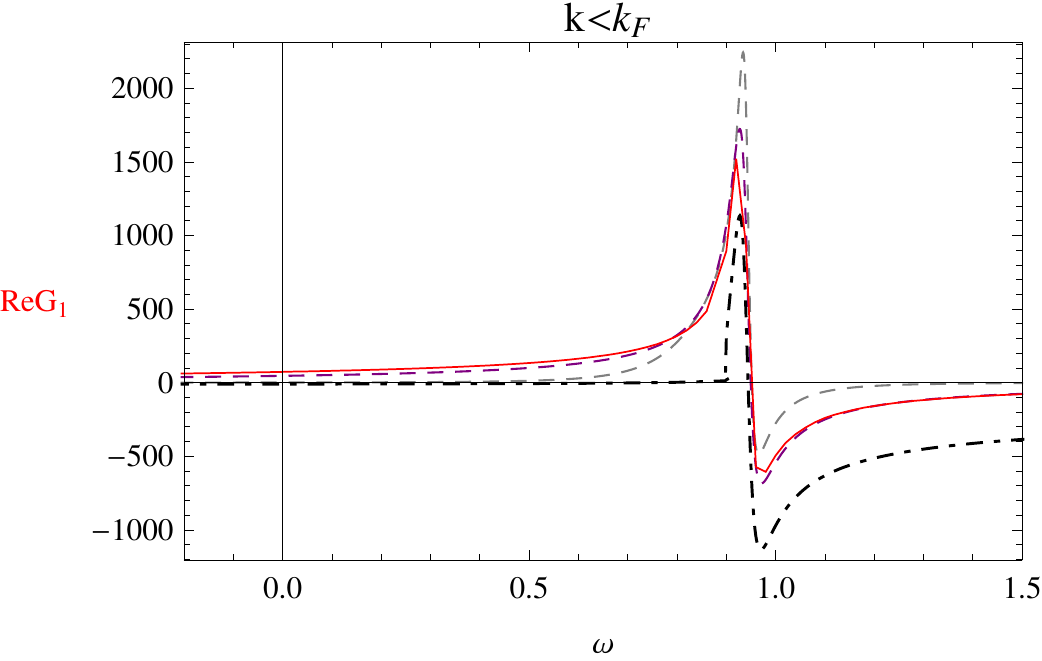, width=3.6 in}} 
\caption{$k=621/500<k_F$, both (a)(b) with three fitting curves: (1)LFL Eq.(\ref{LFL}) with $n=2$ in gray-dashed, (2)LFL Eq.(\ref{LFL}) with $n=1$ in purple-dashed, (3)scaling ansatz form Eq.(\ref{eq:fit1}) in black-dotted-dashed. $c_0, c_1, \gamma_0$ are chosen to be positive but their values on two sides $\omega>\omega_F$ and $\omega< \omega_F$ are chosen differently.} 
\label{fig:fit1} 
\end{figure}

\begin{figure}[!h]
\centerline{ \epsfig{figure=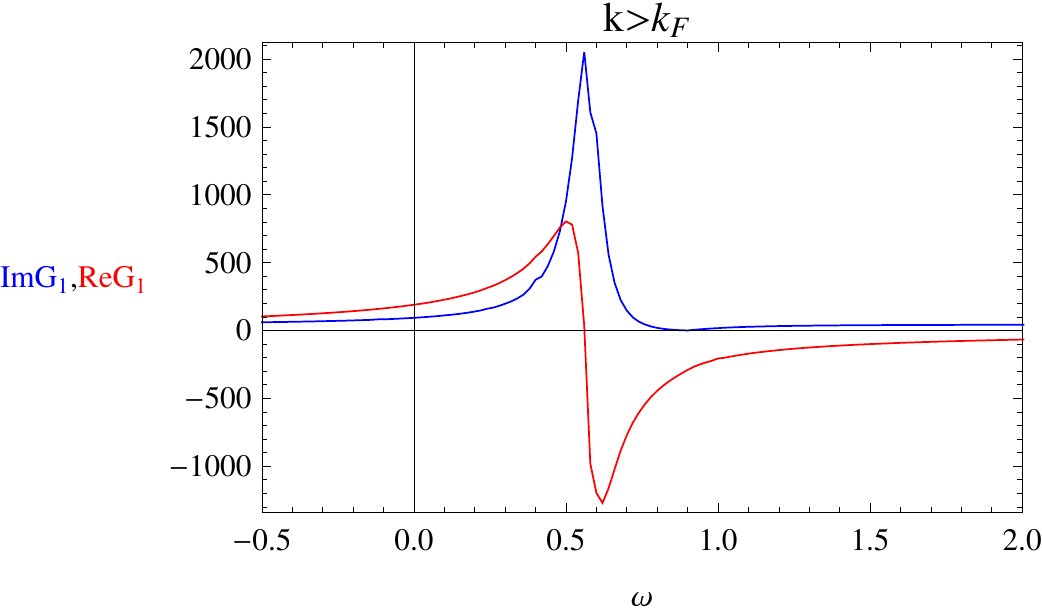, width=3.2 in}} 
\caption{$k=8/5>k_F$. Blue curve is for $\Img G_1$, red curve is for $\Rea G_1$.} 
\label{fig:data2} 
\end{figure}

\begin{figure}[!h]
\centerline{ (a)\epsfig{figure=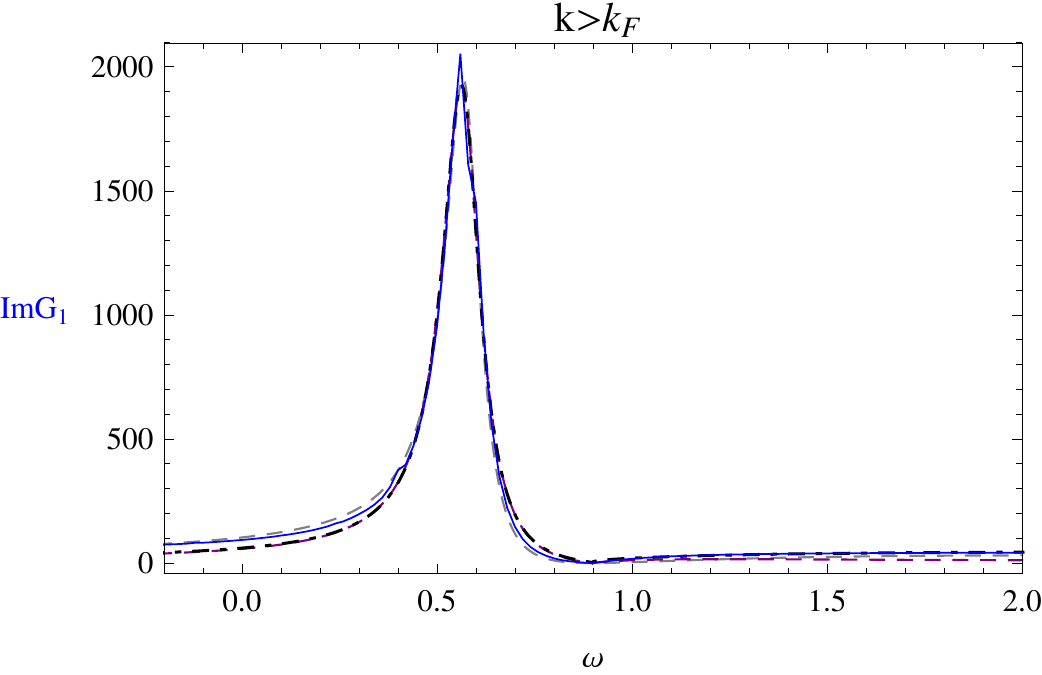, width=3.6 in} (b)\epsfig{figure=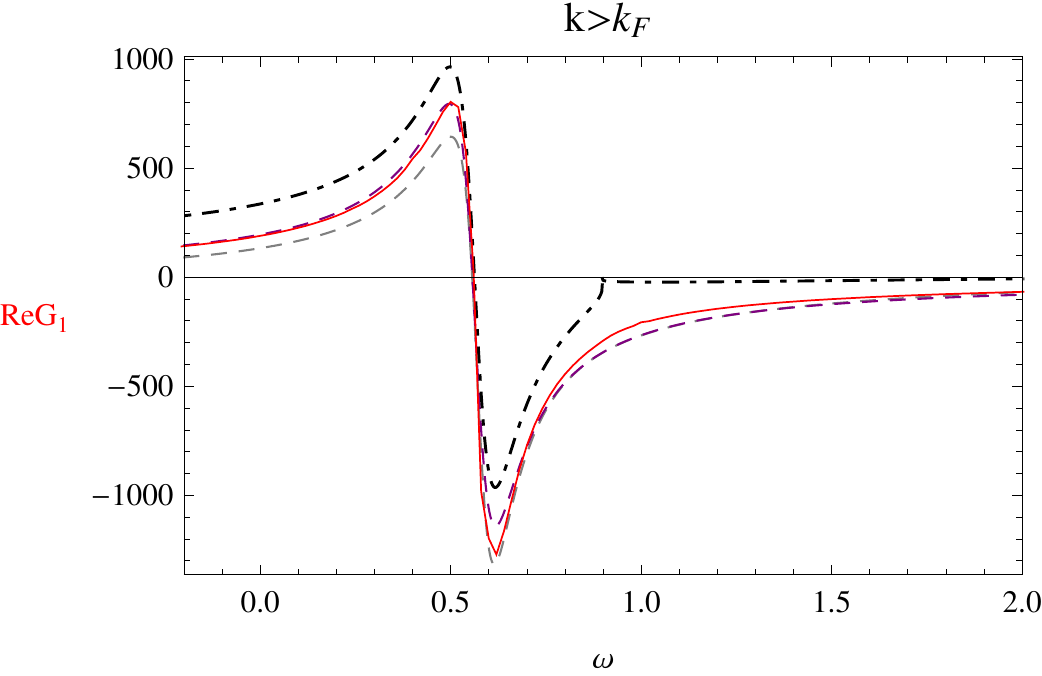, width=3.6 in}} 
\caption{$k=8/5>k_F$, both (a)(b) with three fitting curves: (1)LFL Eq.(\ref{LFL}) with $n=2$ in gray-dashed, (2)LFL Eq.(\ref{LFL}) with $n=1$ in purple-dashed, (3)scaling ansatz form Eq.(\ref{eq:fit2}) in black-dotted-dashed. $c_0, c_1, \gamma_0$ are chosen to be positive but their values on two sides $\omega>\omega_F$ and $\omega< \omega_F$ are chosen differently.} 
\label{fig:fit2} 
\end{figure}

\begin{figure}[!h]
\centerline{ \epsfig{figure=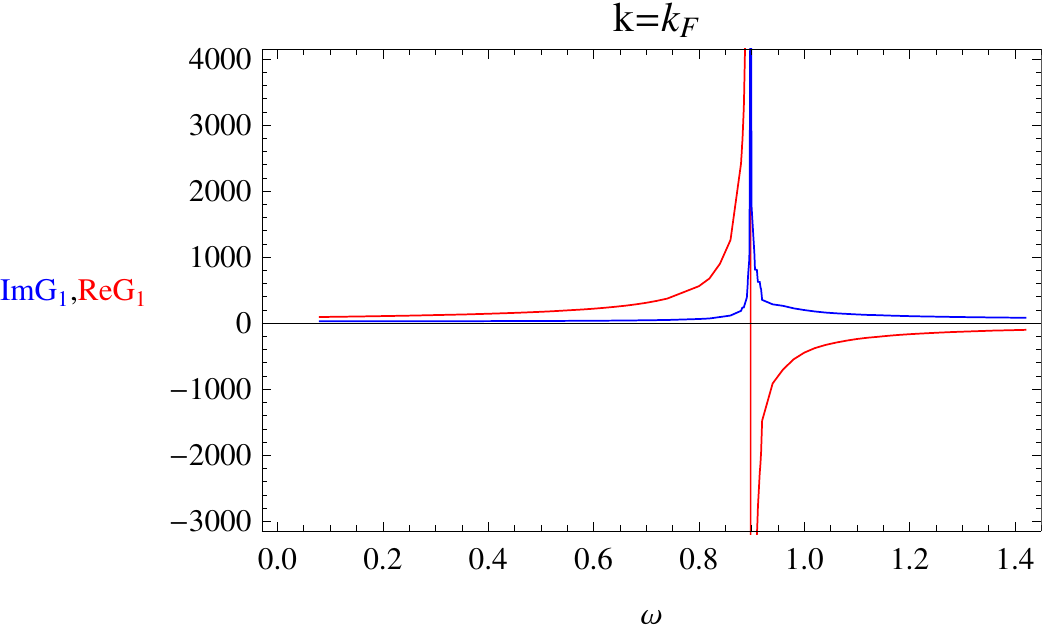, width=3.2 in}} 
\caption{$k=k_F$. Blue curve is for $\Img G_1$, red curve is for $\Rea G_1$.} 
\label{fig:data3} 
\end{figure}

\begin{figure}[!h]
\centerline{ (a)\epsfig{figure=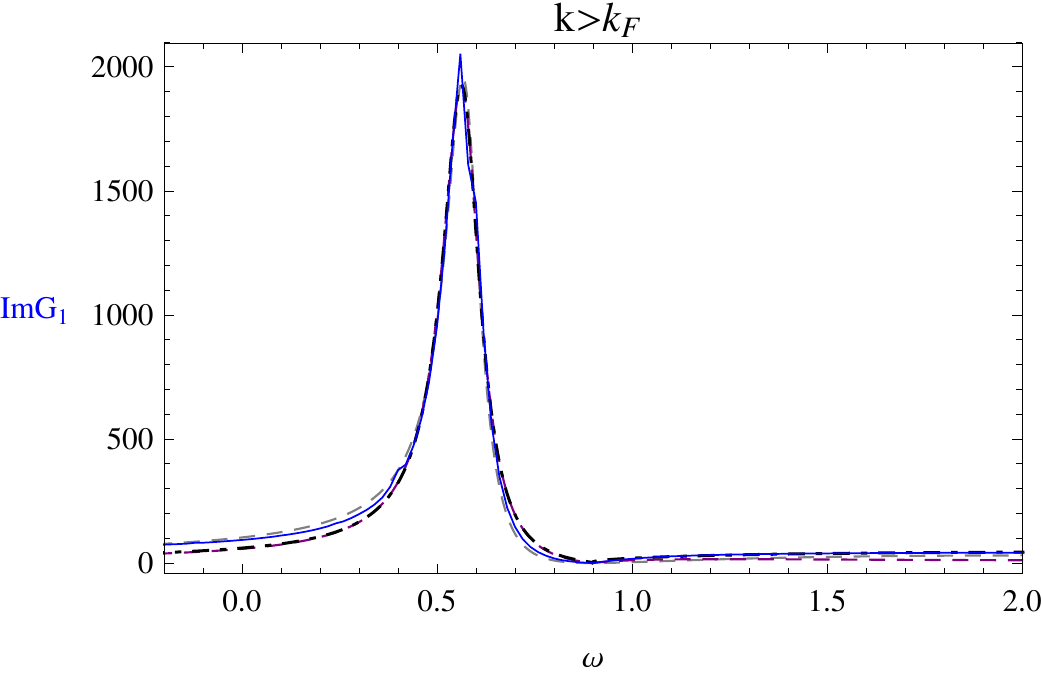, width=3.6 in} (b)\epsfig{figure=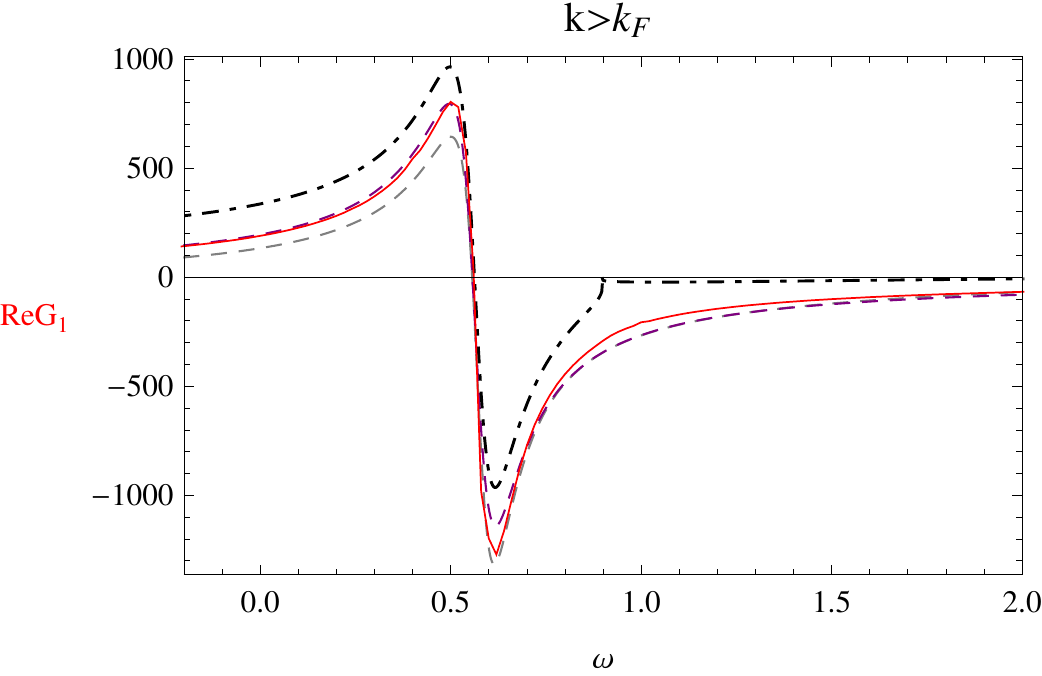, width=3.6 in}} 
\caption{$k=k_F$, both (a)(b) with two fitting curves: (1)LFL Eq.(\ref{LFL}) with $n=2$ in gray-dashed, (2)LFL Eq.(\ref{LFL}) with $n=1$ in purple-dashed. At $k=k_F$, scaling ansatz of both forms Eq.(\ref{eq:fit1}) and Eq.(\ref{eq:fit2}) runs into trouble. We have not yet found a good fitting.} 
\label{fig:fit3} 
\end{figure}

  In the following subsections, we present our Green's function data for $k<k_F$, $k=k_F$ and $k\simeq k_F$, and fit these data by LFL and Sentil's ansatz. 
The main messages of our fitting(\figref{fig:fit1}, \figref{fig:fit2}, \figref{fig:fit3}) of LFL and Sentil's ansatz to our data(\figref{fig:data1}, \figref{fig:data2}, \figref{fig:data3}) are: \\
(1) Sentil's ansatz generally has better agreement than LFL fitting for our $\Img G_1$ data.\\
(2) Our $\Rea G$ data are sandwiched by the LFL fitting with LFL fitting with $n=2$ and Marginal Fermi liquid(MFL) with $n=1$\cite{Varma:1989zz,Iqbal:2011ae}, 
which likely implies that our Schr\"odinger Fermi liquids can be a closer description between LFL and MFL theory with $1<n<2$. 
From our quasiparticle self-energy ansatz as $\gamma(\omega)=\kappa(\omega-\omega_F)^n$ and quasiparticle decay rate ${1}/{\tau_k}\sim \gamma(\omega)$,
this may suggest Schr\"odinger Fermi liquids has shorter life time and larger decay rate ${1}/{\tau_k} \sim (\omega-\omega_F)^{2-\varepsilon}$ than LFL 
${1}/{\tau_k} \sim (\omega-\omega_F)^{2}$ close to Fermi energy $\omega_F$. Compared to LFL, the quasiparticle description of Schr\"odinger Fermi liquids is less robust.\\
(3) Near the pole location, we have not found a promising fitting for Sentil's ansatz for both sides of $\omega>\omega_F$ and $\omega< \omega_F$.

\newpage
\subsubsection{$k<k_F$}

For $k<k_F$, our scaling ansatz is
\footnote{
For $\omega<\omega_F$ under $k<k_F$, the term inside logarithmic becomes negative, which we choose the complex logarithm as following, 
\be
\frac{c_0(k-k_F)^{-\alpha}}{\log(\frac{(\omega-\omega_F)}{c_1(k-k_F)^z})+i \pi +i \gamma_0}
\ee}
\be \label{eq:fit1}
\frac{c_0(k-k_F)^{-\alpha}}{\log(\frac{-(\omega-\omega_F)}{c_1(k-k_F)^z})+i \gamma_0}
\ee

\newpage
\subsubsection{$k>k_F$}

For $k>k_F$, our scaling ansatz is 
{\footnote{
For $\omega>\omega_F$ under $k>k_F$, the term inside logarithmic becomes negative, which we choose the complex logarithm as following, 
\be
\frac{c_0(k-k_F)^{-\alpha}}{\log(\frac{(\omega-\omega_F)}{c_1(k-k_F)^z})-i \pi -i \gamma_0}
\ee}}
\be \label{eq:fit2}
\frac{c_0(k-k_F)^{-\alpha}}{\log(\frac{-(\omega-\omega_F)}{c_1(k-k_F)^z})-i \gamma_0}
\ee

\newpage
\subsubsection{$k=k_F$.}

LFL fitting must be symmetric respect to $\omega=\omega_F$ at $k=k_F$ for $\Img G_1$, however the $\Img G_1$ from Schr\"odinger Fermi liquids is not symmetric along $\omega=\omega_F$. This is the major difference.

\pagebreak
\newpage

\section{Fermionic Quantum Phase Transition} \label{sec5}
Schr\"odinger black hole introduce two additional parameters to the non-relativistic conformal background, 
other than the parameters occurred already in asymptotic AdS spacetime\cite{Liu:2009dm}. 
The first parameter is the mass operator $M$(particle number eigenvalue), 
the second one is background density $\beta$\footnote{
Usually a quantum phase transition is tuned by a dimensionless coupling\cite{Sachdev:QPT}, in our case we can define 
the coupling $g_\beta\equiv\beta\sqrt{\mu_Q}$ with $\mu_Q$ fixed in our case. 
Since the Hamiltonian description of boundary theory is unknown,
we schematically tune $\beta$ to address the same physics as tuning $g_\beta$.
}. 
In the zero temperature phase, a natural question arises: what happened to the Fermi surface if background density $\beta$ is tuned? We should fix $M$ while varying $\beta$
\footnote{Here we tune $\beta$ with various values $1/16,1/4,1/2,1/\sqrt{2},1,2,8$. The other parameters should be fixed. We choose $M=\ell+qQ\beta$ fixed to be $1/10$, $T=0(Q=\sqrt{2})$, $\Delta_\pm(\nu_\pm)$ is fixed by $m=1/10$ and $M$. Among all the five parameters of the system, the remained parameters $\mu_Q$ is subtle, which is $\mu_Q= Q/(2\beta)$. In our numerics, we choose to fix $q=1$, in this case it seems like chemical potential $\mu_Q$ varies while $\beta$ is tuned. One may argue that a resolution is considering $\mu_q\equiv q \mu_Q$ where $\mu_q$ still allowed to be fixed while $q$ compensates to be adjusted correspondingly. This resolution seems to fix the (chemical potential) energy to add a fermion of charge q into the system. However, we should aware that in any case the chemical potential $\mu_Q$ is indeed varied. In addition, the `real' chemical potential to set the scale of Fermi energy $\mu_F$ is not merely as in \cite{Liu:2009dm} only $\mu_q$ alone. In our Schr\"odinger system, the Fermi energy should be identified by the coordinate $\tilde{\omega}$, the Fermi energy $\mu_F$ is set by $\tilde{\omega}+\mu_F=\big(\beta(\omega+q A_t)+(l-q A_\xi)/(2 \beta) \big)\vert_\partial=\tilde{\omega}$. This shows that $\mu_F=q \big(\beta A_t- A_\xi/(2 \beta) \big)\vert_ \partial =0$ is independent of $\beta$. Therefore, Fermi energy $\mu_F=0$ is already fixed. We choose to fix the charge $q$ of fermionic contents, instead of varying fermion charge $q$ to fix the energy $\mu_q$ of inserting a fermion}. 
In Sec.\ref{sec:beta>*}, we encounter again the phase with Fermi surfaces, gradually tune down $\beta$ near $1/2$, we find a critical point(or critical line) in Sec.\ref{sec:beta=*}. Smaller $\beta$ shows that Fermi surface collapses and then disappears in Sec.\ref{sec:beta<*}. 
Altogether may indicate a quantum phase transition of fermionic liquids.

\subsection{Well-defined Fermi surface ($\beta>\beta^*$)} \label{sec:beta>*}
When $\beta > 1/2$, an obvious peak appears in $\Img G_1$, see \figref{fig:>}. Analytically the peak should approach $\delta(k-k_F, \omega-\omega_F)$ at $(k_F,\omega_F)$. 
However, the numerical value $\Img G_1(k_F,\omega_F)$ cannot really be infinite. 
What we find is that the peak $\Img G_1(k_F,\omega_F)$ values in this region $\beta > 1/2$ depends on the finite IR cutoff. 
The smaller the initial cutoff $\epsilon=r-1$, the larger the numerics $\Img G_1(k_F,\omega_F)$ at the peak grows. This is a sign for the suppose-to-be infinite pole.
The pole of $\Img G_1$ indicates a well-defined Fermi surface.
At larger $\beta$, the pole and nearby region on $(k,\omega)$ plane develops much sharper. 
Notice that the pole shifts to larger $\omega_F$ and smaller $k_F$ by decreasing $\beta$.

\subsection{Near the quantum critical point  ($\beta \simeq\beta^*$)} \label{sec:beta=*}
As $\beta$ approaches in the range between $1/\sqrt{2}$ and $1/2$, 
we find the $\Img G_1(k_F,\omega_F)$ peak becomes insensitive to IR finite cutoff $\epsilon$, the peak values are lower for smaller $\beta$, shown in \figref{fig:=}. 
The stable peak value indicates there is no $\delta$-function like pole on $(k,\omega)$ plane.  By tuning $\beta$ to smaller value, the Fermi surface gradually collapse and disappear. 
We interpret the physics as: 
\bea \label{Z}
\beta > 1/2,\;\;\; \Img G_1(k_F,\omega_F) \simeq Z\delta(k-k_F, \omega-\omega_F)+ \text{finite terms},\;\;\; \text{where } Z\neq0.\\
\beta \simeq 1/2,\;\;\; \Img G_1(k_F,\omega_F) \simeq Z\delta(k-k_F, \omega-\omega_F)+ \text{finite terms},\;\;\; \text{with } Z\rightarrow 0
\eea
Since the $Z$ goes to zero at finite $\beta \simeq 1/2$, we suspect it is not a smooth crossover behavior.
We expect a quantum critical point $\beta^*$(or quantum critical line) slightly larger than $\beta=1/2$, and smaller than $\beta=1/\sqrt{2}$. 
For convenience, we denote  $\beta^*\simeq 1/2$, as in \figref{fig:=}(b).
We do not numerically determine $\beta^*$ due to the computational limitation.
A more detailed scan near the peak at $\beta \simeq 1/2$ may determine the exact value of $\beta^*$.

\subsection{Fermi surface collapse and disappearance ($\beta<\beta^*$)} \label{sec:beta<*}
\begin{figure}[!h]
\centerline{ (a)\epsfig{figure=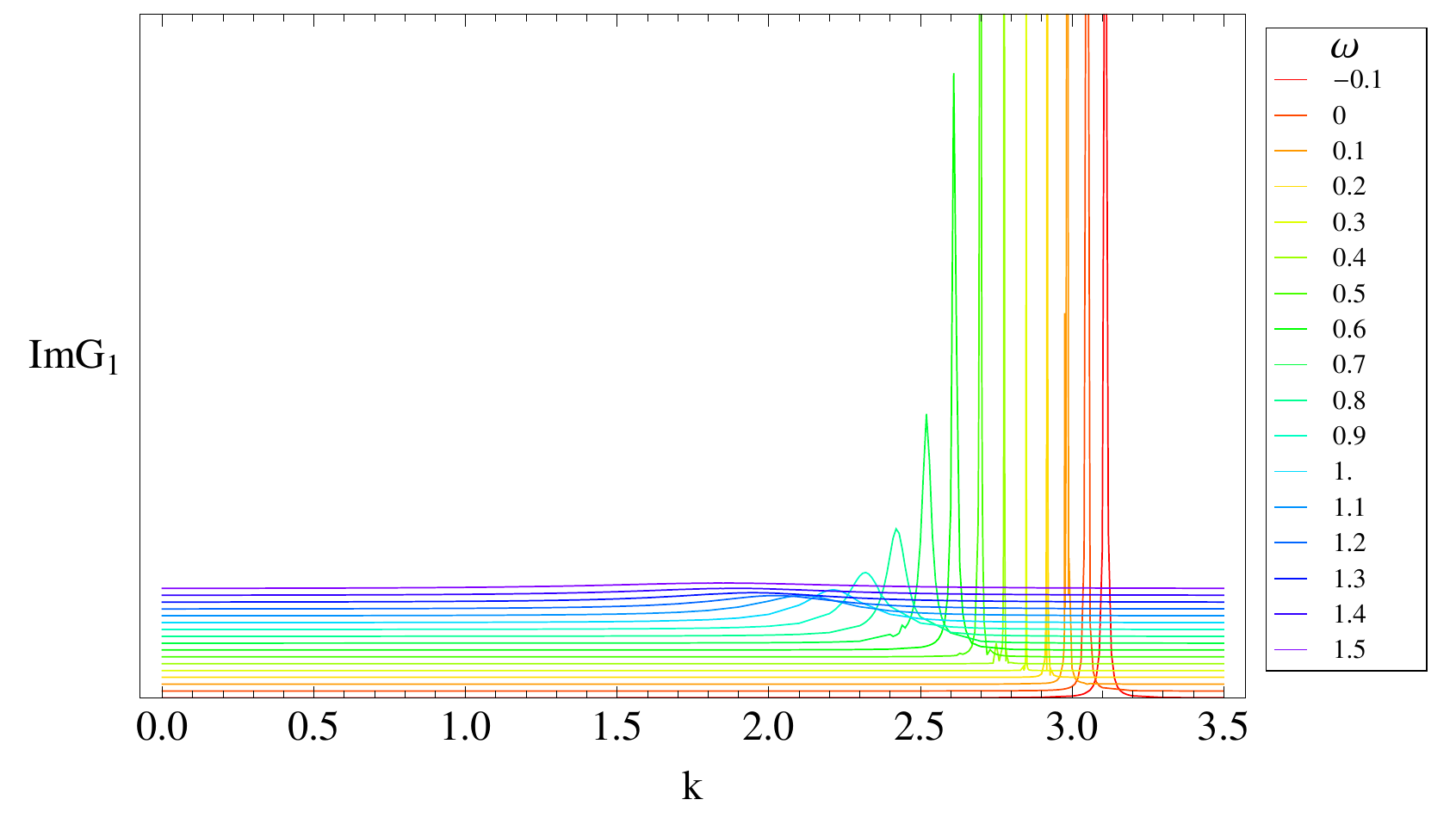, width=3.6 in} (b)\epsfig{figure=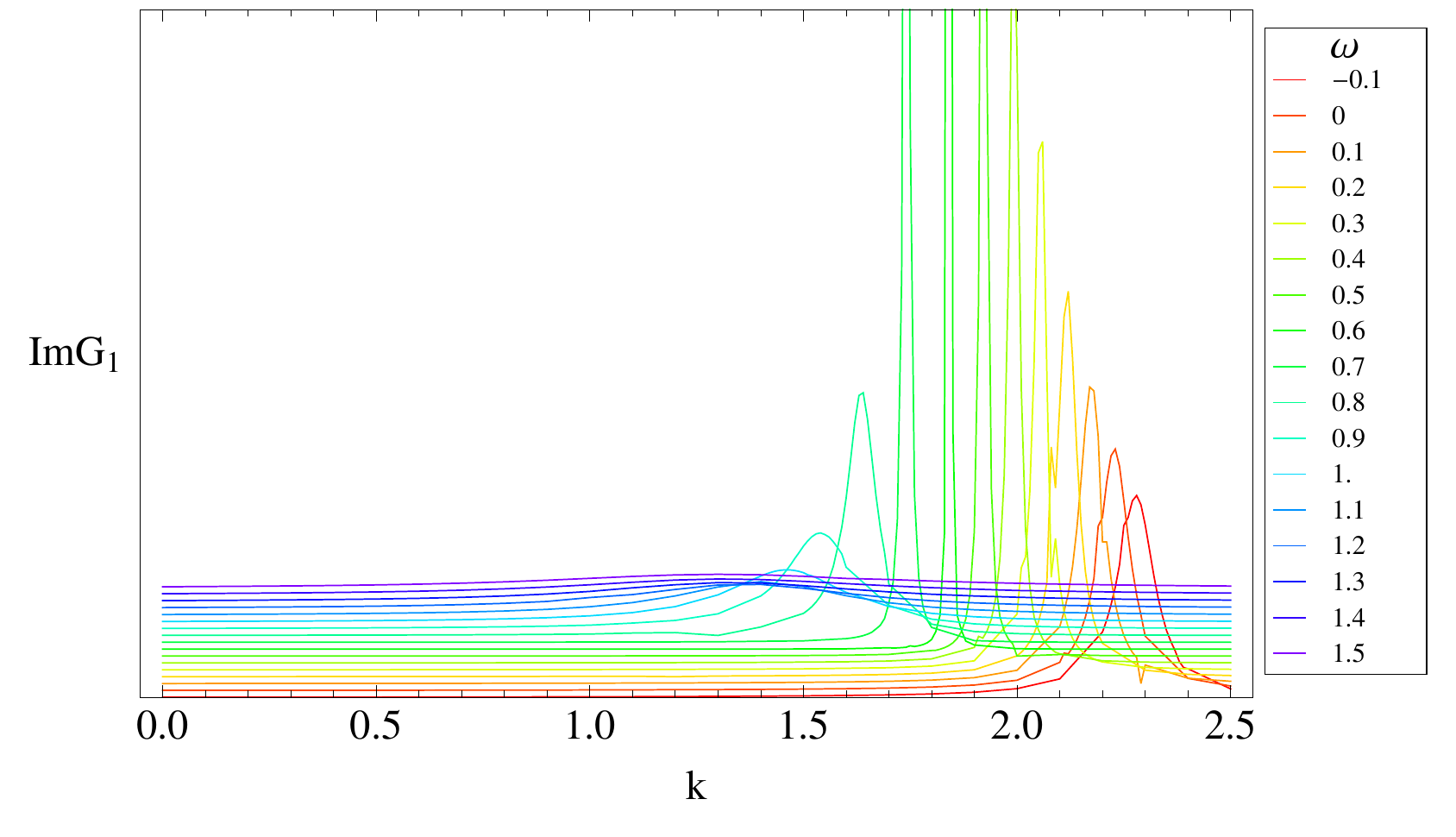, width=3.6 in}} 
\caption{$\Img G_1$ of Schr\"odinger Fermi liquids for $\beta>\beta^*$: (a)$\beta=2$\;\; (b)$\beta=1$} 
\label{fig:>} 
\end{figure}

\begin{figure}[!h]
\centerline{ (a)\epsfig{figure=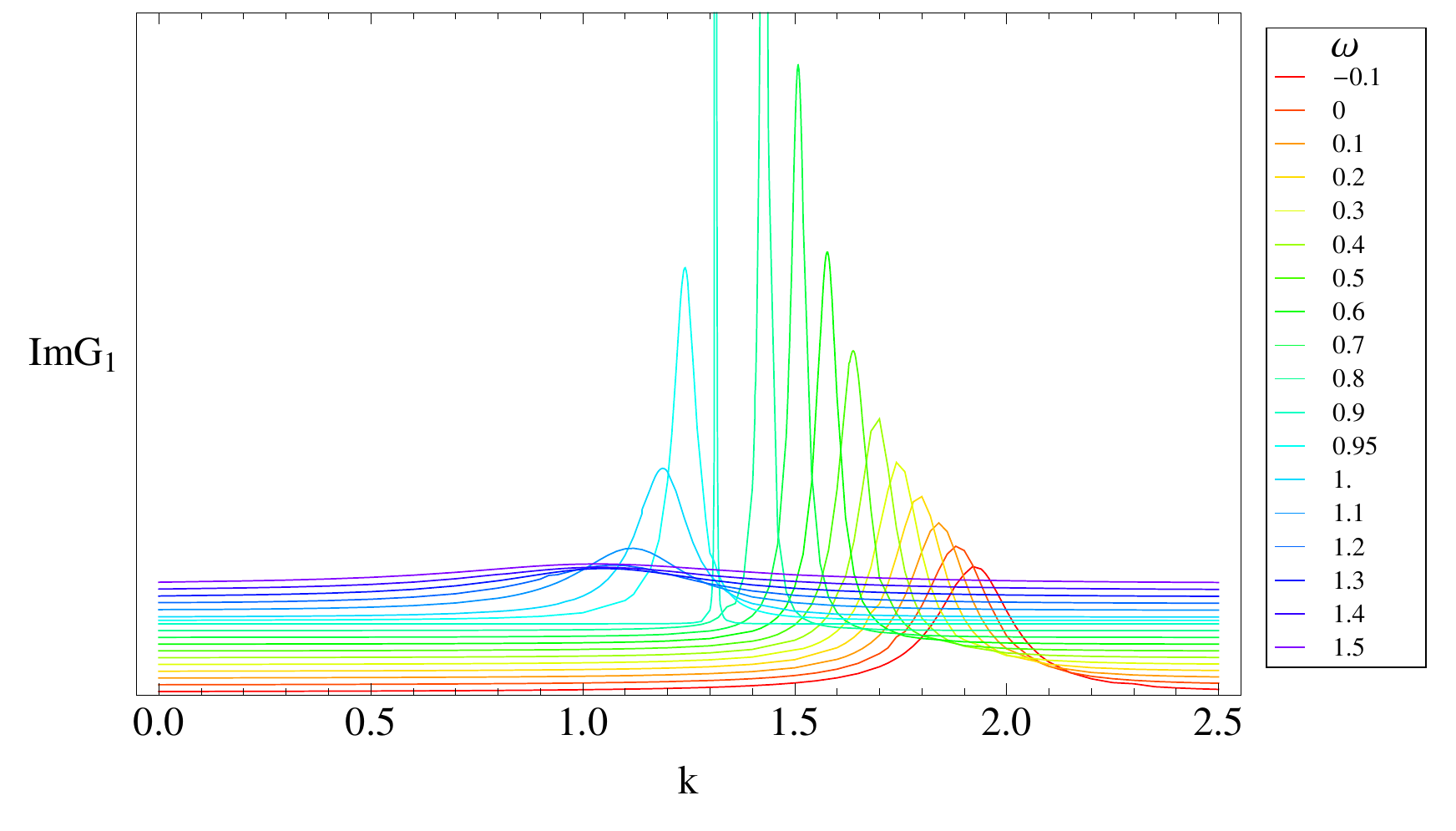, width=3.6 in}  (b)\epsfig{figure=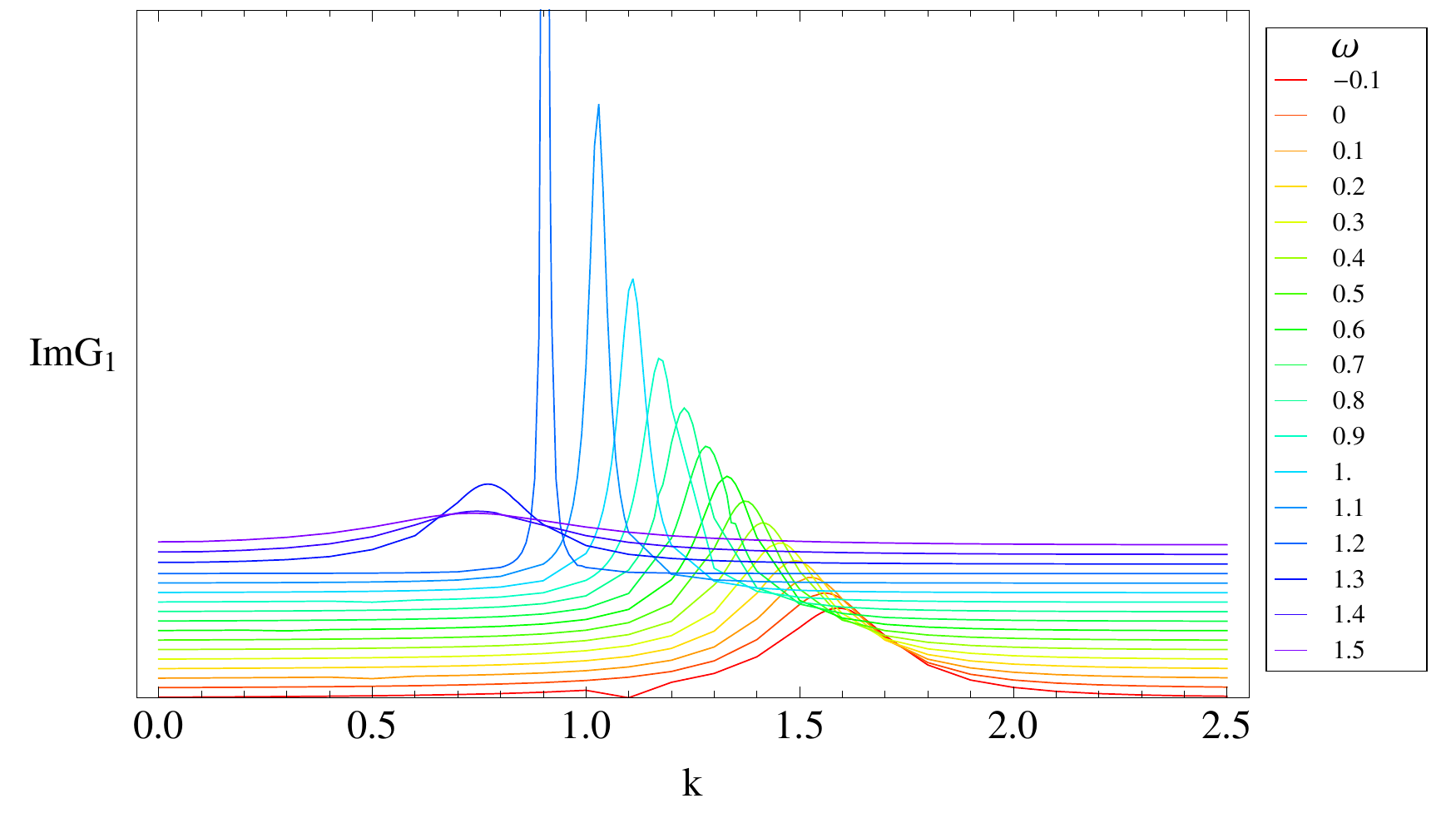, width=3.6 in} } 
\caption{$\Img G_1$ of Schr\"odinger Fermi liquids: (a)$\beta=1/\sqrt{2}$ (b)$\beta=1/2\simeq \beta^*$. Near $\beta^*$.} 
\label{fig:=} 
\end{figure}

\begin{figure}[!h]
\centerline{(a)\epsfig{figure=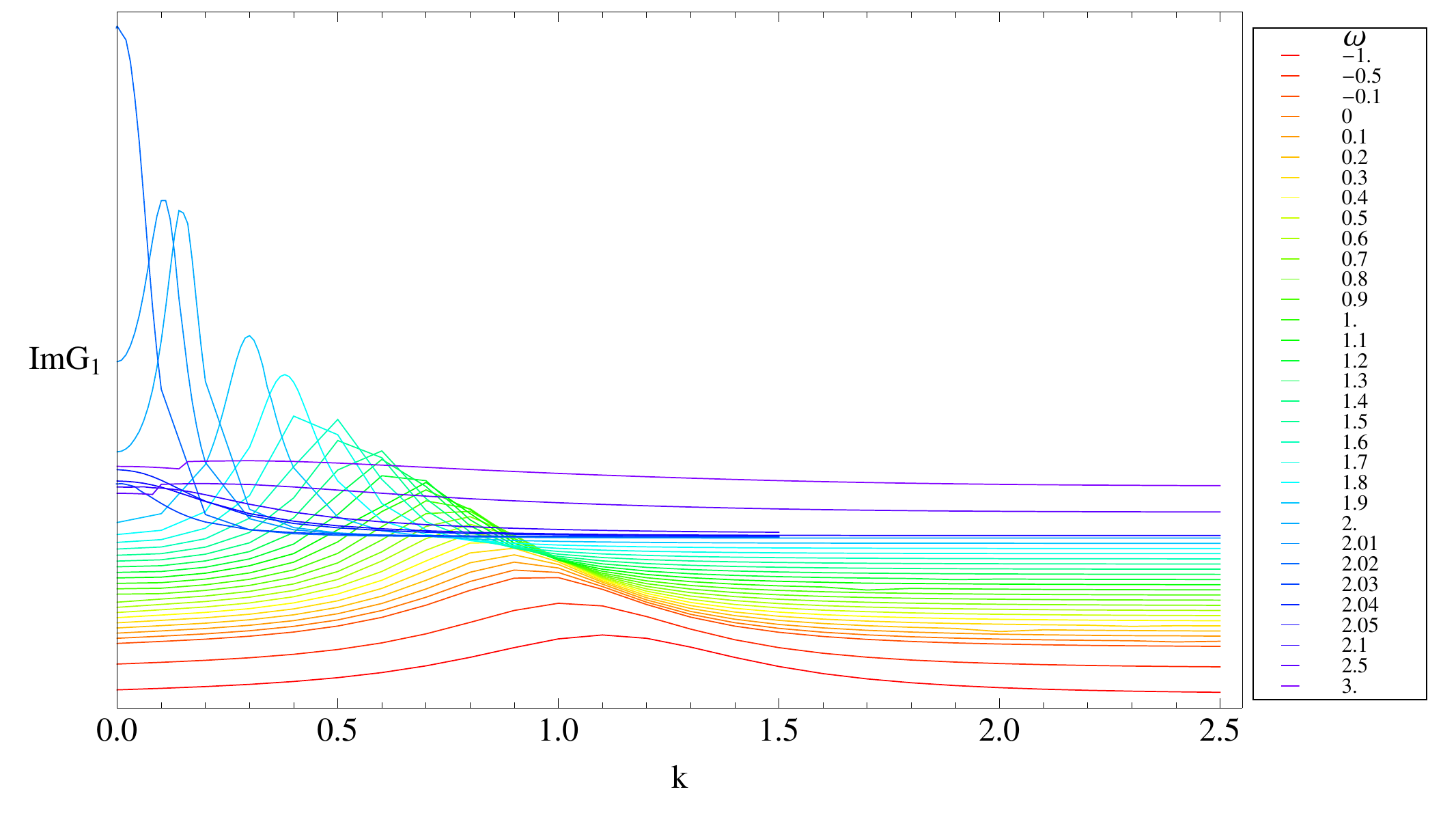, width=3.6 in} (b)\epsfig{figure=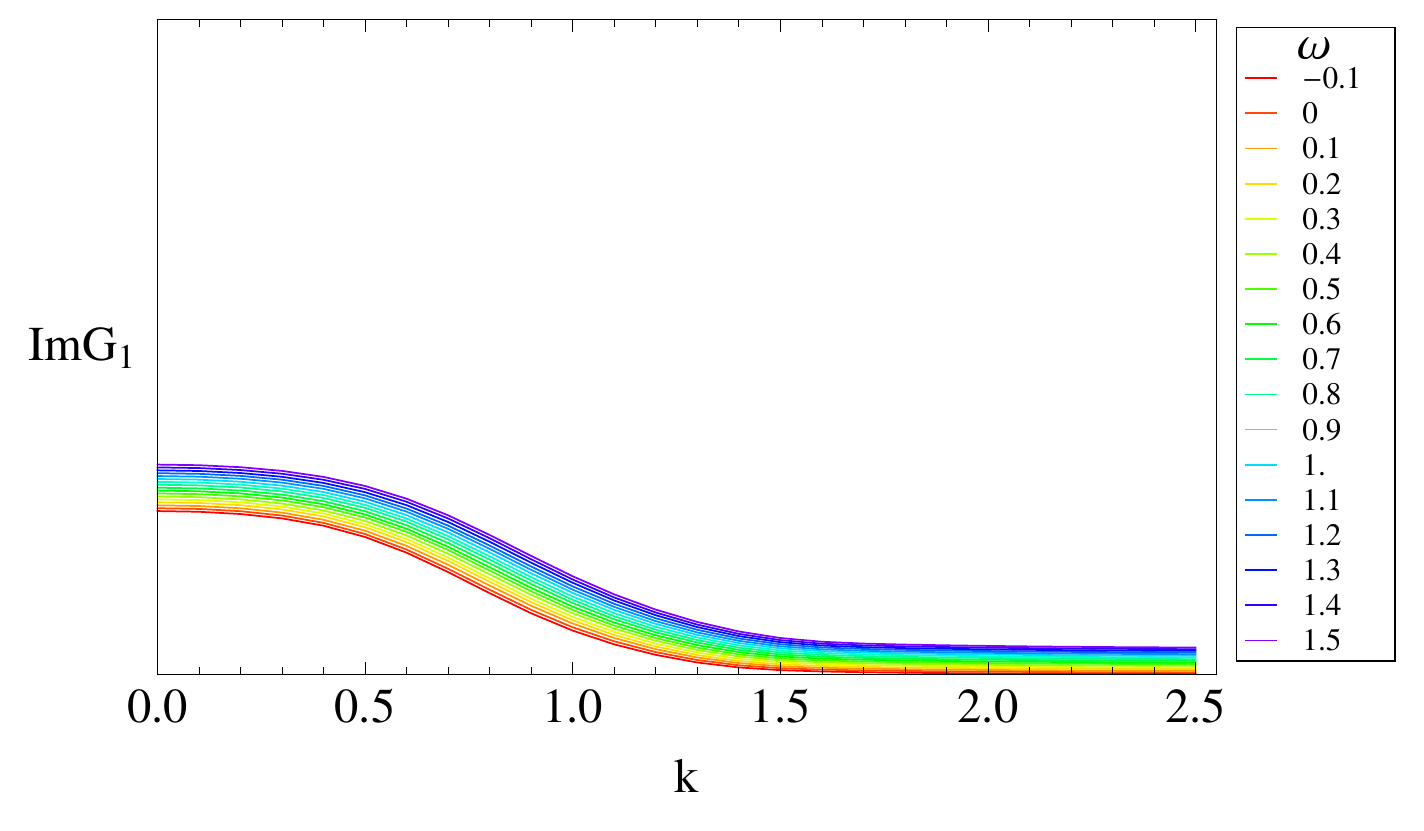, width=3.6 in}} 
\caption{$\Img G_1$ of Schr\"odinger Fermi liquids for $\beta<\beta^*$: (a)$\beta=1/4$\;\; (b)$\beta=1/16$} 
\label{fig:<} 
\end{figure}

At smaller $\beta < 1/2$, we find no sharp peak but only a smoother hump. Unlike Mott insulator for Mott transition \cite{Senthil,senthil:0804}, we do not have gaps opened up in $A(k,\omega)$.
Indeed $A(k,\omega)$ does not dip to zero in this phase. There is no non-analiticity in $A(k,\omega)$ to pin point $k_F$. This shows it is still a gapless phase but with Fermi surface disappearance.
We show a series of $\Img G_1$ plots by varying $\beta$ in Fig.\ref{fig:>},\ref{fig:=},\ref{fig:<}. 
Note that the vertical axes for $\Img G_1$ shows no tick marks, we only show a landscape scanning through many slices of $\Img G_1$.
Each slice of $\Img G_1(k,\omega)$ has a fixed $\omega$, and scanning  $k$ values.
Each slice of $\Img G_1$ has been shifted vertically for a clear vision of the landscape. 
As in Footnote.\ref{footnote:akw} we had discussed the $L_\xi$ size of the compact $\xi$ circle modifies $\Img G_1$ to the physical value of $A(k,\omega)$. 
Therefore, here the exact value of $\Img G_1$ is immaterial, only the relative height of $\Img G_1$ matters.


\newpage
\subsection{Evolutions of dynamical exponent $z$ and Fermi-momentum $k_F$ under tuning background density $\beta$} 

\begin{figure}[!h]
\centerline{(a) \epsfig{figure=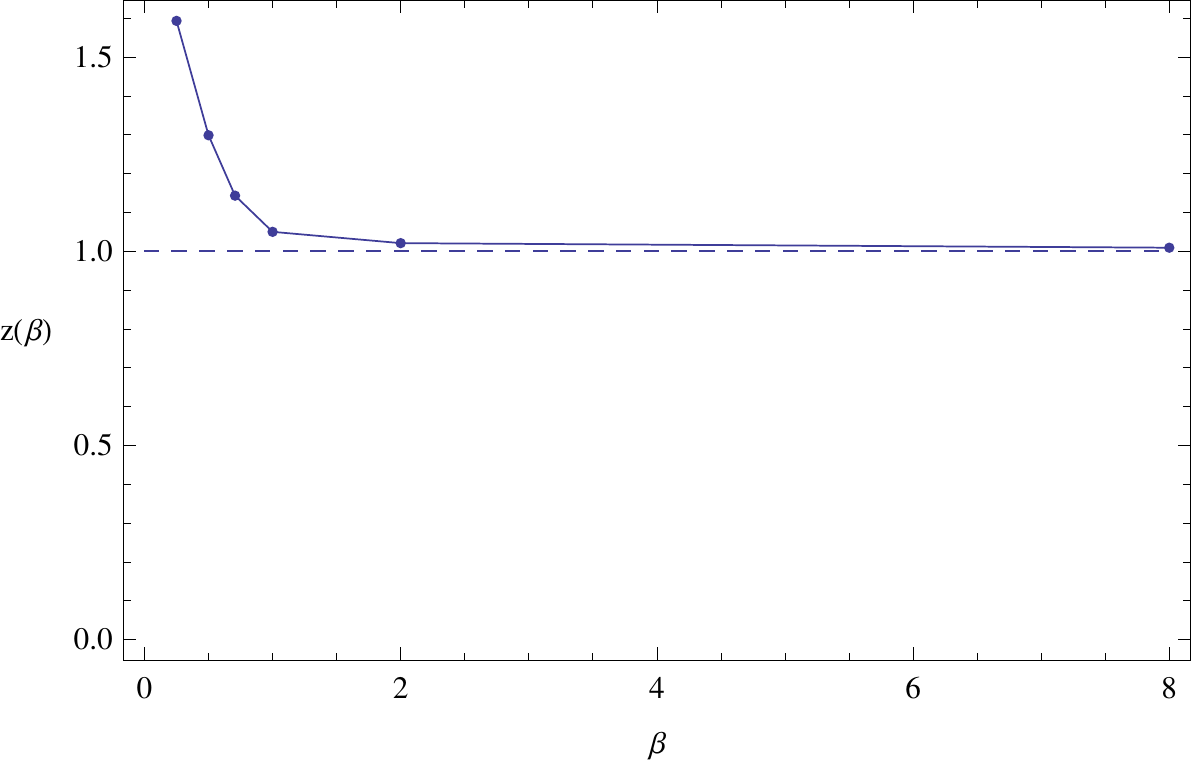, width=3.56 in}   (b) \epsfig{figure=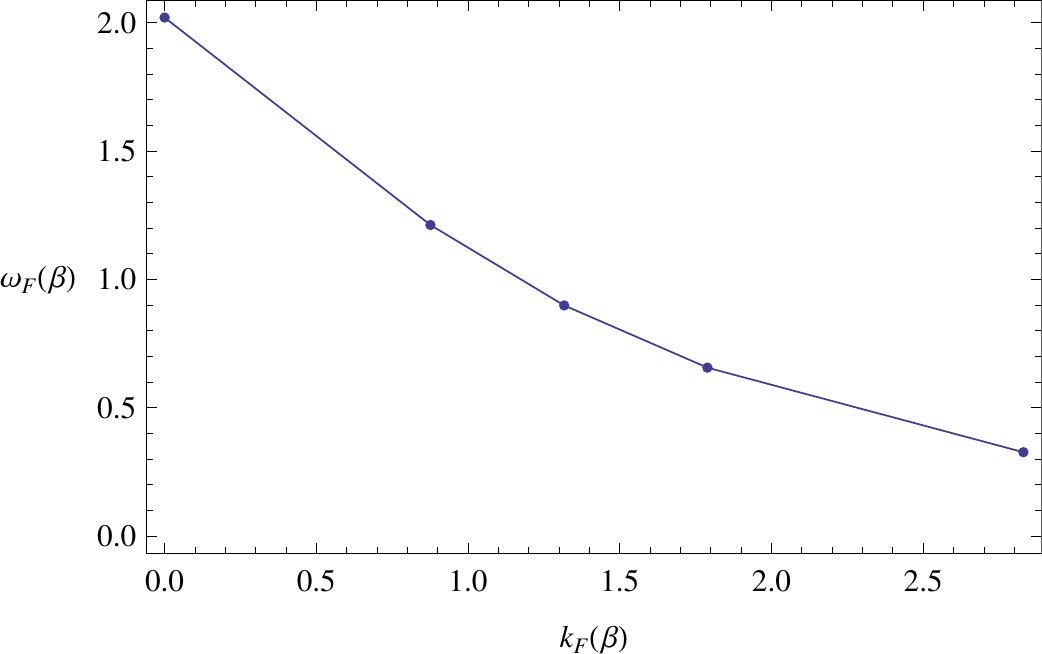, width=3.6 in}} 
\caption{(a)The relation between $z$ and $\beta$ for six data points are: 
$\beta=1/4, z=1.593$, $\beta=1/2,  z=1.299$, $\beta=1/\sqrt{2}, z=1.143$, $\beta=1, z=1.050$, $\beta=2, z=1.021$, $\beta=8, z=1.009$.
(b)For various $\beta=1/4,1/2,1/\sqrt{2},1,2$, the Fermi-momentum and energy $(k_F,\omega_F)$ are $(0.0000303, 2.020)$, 
$(0.8762, 1.212)$, $(1.317, 0.8984)$, $(1.789, 0.656)$, $(2.8304, 0.3266)$.} 
\label{z_vs_beta} 
\end{figure}

\begin{figure}[!h]
\centerline{(a)\epsfig{figure=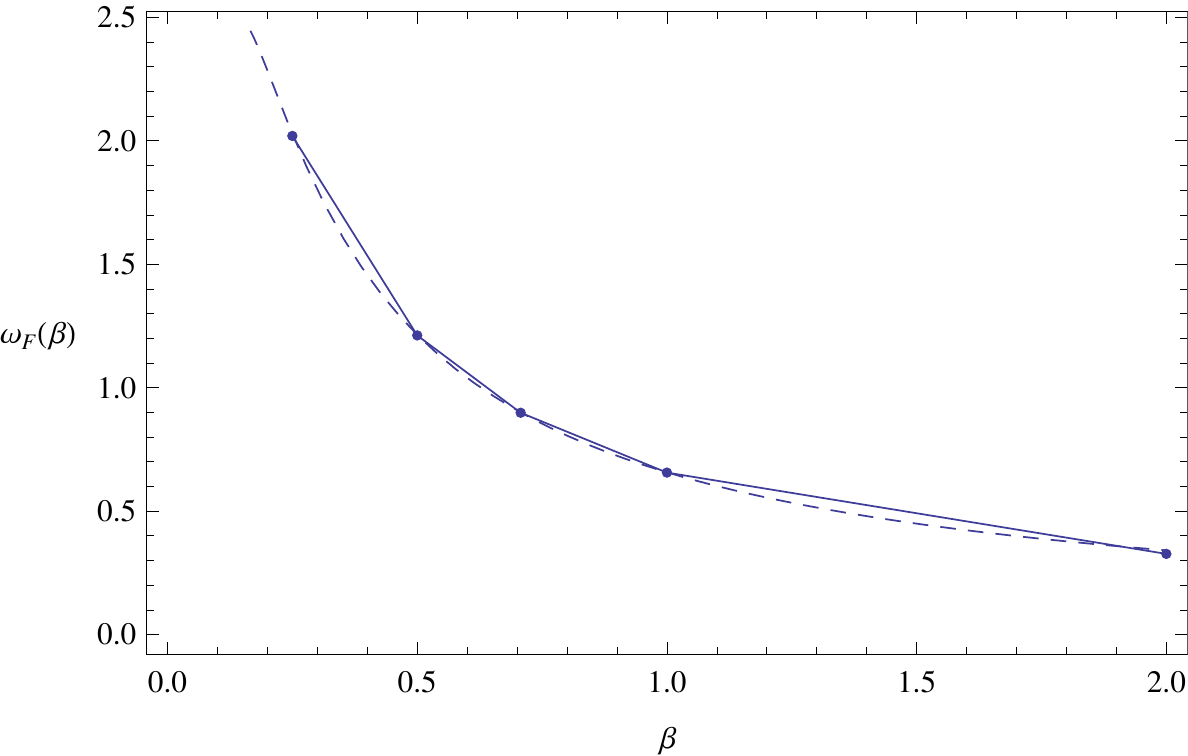, width=3.6 in} (b)\epsfig{figure=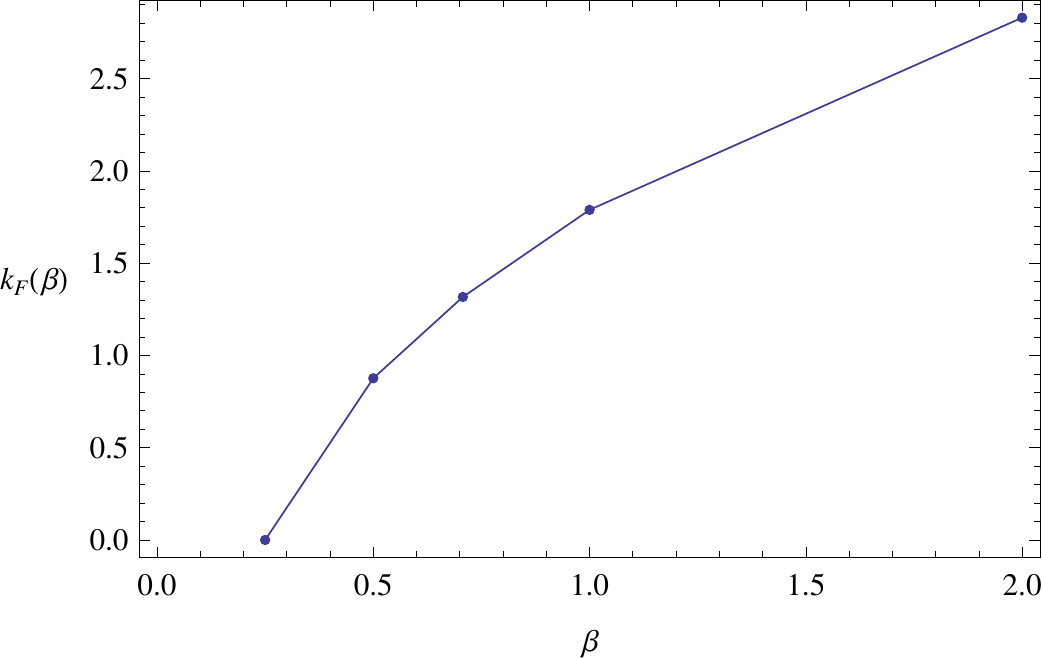, width=3.56 in} } 
\caption{ (a)The location of the $\Img [G_1]$ peak at $\omega_F$ varies respect to $\beta$. The numerical data points compared to the dashed curve fitting $\omega_F(\beta)=-(M-qQ\beta)/(2\beta^2)$.
(b)The location of the Fermi-momentum $k_F$ varies respect to $\beta$.
} 
\label{w_vs_beta} 
\end{figure}

Here we study the evolutions of the dynamical exponent $z$,  Fermi energy $\omega_F$, Fermi-momentum $k_F$ while tuning $\beta$ in \figref{z_vs_beta}, \figref{w_vs_beta}. 
When $\beta \lesssim \beta^*$, there is no good quasiparticle description for the system, 
so in which case $(\omega_F,k_F)$ means the $(\omega,k)$ coordinates of the highest peak in spectral function,
$z$ just means the dispersion reading from the branches structure around the highest peak.

In all of the data above, $\alpha \simeq 1.00$, thus our data follows the general relation $z \geq \alpha$ and $z \geq 1$ as Senthil's argument\cite{Senthil,senthil:0804}. 
As $\beta$ increases, $z$ goes close to 1. Tentatively it suggest a more Landau Fermi liquids like behavior at large $\beta$ limit. Though from the spectral density fitting, we find the imaginary part of quasiparticle does not obey $\gamma(\omega) \propto \omega^2$ and $\Img [G_1(k,\omega)]$ near $k_F$ is not symmetric respect to $\omega_F$. These two features are distinct from LFL. The large $\beta$ limit is at most a close cousin of LFL.\\

In Sec(\ref{sec:FS}), we showed $\omega_F=-\ell/(2\beta^2)$. In the case of tuning $\beta$ while fixing the mass operator $\ell-q M_{o} =\ell+qQ\beta=M$, we expect $\omega_F(\beta)=-(M-qQ\beta)/(2\beta^2)$. We show this power law fitting agrees with our data in \figref{w_vs_beta}(a). On the other hand, the Fermi-momentum $k_F$ requires better understanding of UV physics, we do not have a fitting here.

\pagebreak

\newpage

\section{Conclusion and Open Questions} \label{sec6}

\begin{figure}[!h]
\centerline{\epsfig{figure=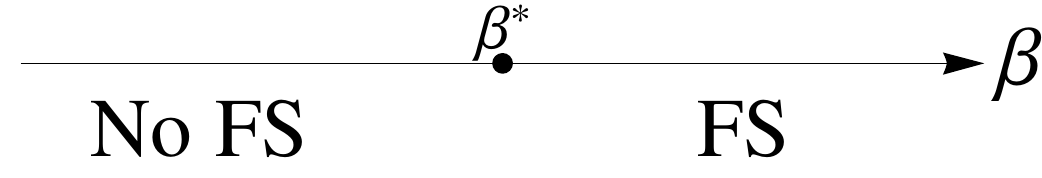,width=3. in }} 
\caption{The zero T quantum phase transition diagram of Schr\"odinger Fermi liquids. 
A phase with Fermi surface(FS) in $\beta > \beta^*$.
A quantum critical region near $\beta \simeq \beta^*$
(it is undetermined yet whether $ \beta^*$ is a critical point or critical line, and unknown whether the transition is 1-st order or higher order). 
A phase without Fermi surface in $\beta< \beta^*$. 
Note to tune a dimensionless coupling $g_\beta$, we can define $g_\beta\equiv\beta\sqrt{\mu_Q}$ with $\mu_Q$ fixed.
} 
\label{fQPT} 
\end{figure}

In summary, we have studied a class of strongly interacting non-relativistic fermions under asymptotic NRCFT background  in $2+1$ D. 
We make some efforts to deal with the aforementioned two shortcomings of AdS space. 
Firstly, our model has a better realization of non-relativistic properties of many body systems, and we have observed the well-defined Fermi surface of Schr\"odinger Fermi liquids.
Secondly, by tuning the background density $\beta$ with fixed particle number $M$, we realized a fermionic quantum phase transition as \figref{fQPT}, 
where on larger $\beta$ side shows a sharp Fermi surface, 
while on smaller $\beta$ side shows Fermi surface disappearance. 
We find Senthil's scaling ansatz generally a better fit than Landau Fermi liquid(LFL) to our non-Fermi liquids.
Based on quasiparticle self-energy scaling, we argue the quasiparticle description of Schr\"odinger Fermi liquids has shorter lifetime and is less stable comparing to LFL.\\

We leave some questions for future directions:\\
\noindent
(1) Quasi-particle residue $Z$ may be regarded as the order parameter for the quantum phase transition. How does $Z$ in Eq.(\ref{Z}) behave near quantum critical point(or line), what is the order of phase transition? 
We have not yet been able to answer these questions. 
It will be illuminating to understand whether Schr\"odinger Fermi liquids shows 
discontinuous 1st, or continuos 2nd order or higher order transition, and the possibility to realize similar phase transitions proposed in \cite{Senthil,senthil:0804}. 
It is also noteworthy that the location of poles has been captured very well analytically by the speculated curve $\omega_F(\beta)=-(M-qQ\beta)/(2\beta^2)$ as in \figref{w_vs_beta} , though 
we see numerical data slightly deviated from the analytic curve (at 3 digits after the decimal mark). It will be important to know the physical mechanism or subleading corrections for this deviation. \\

\begin{figure}[!h]
\centerline{\epsfig{figure=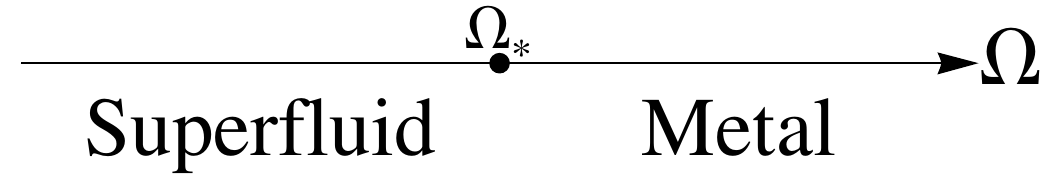,width=3. in }} 
\caption{Bosonic quantum phase transition is likely found at low temperature phase in \cite{Adams:2011kb}. 
On large background density($\Omega>\Omega_*$) side, the phase is in metallic state with unbroken U(1) symmetry. On smaller background density($\Omega < \Omega_*$) side, there shows a superfluid phase with broken U(1) symmetry.
Note to tune a dimensionless coupling $g_\Omega$, we can define $g_\Omega\equiv \Omega /\sqrt{\mu_Q}$ with $\mu_Q$ fixed.
} 
\label{bQPT} 
\end{figure}

\noindent
(2) 
Notably the charge or particle number U(1) symmetry are unbroken in our probe limit.
Fermi surface and gauge-gravity duality relation are mentioned in \cite{Sachdev:2011ze,Huijse:2011hp},
especially the relation between a global U(1) symmetry and the existence of Fermi surface. 
How does our system realize a quantum phase transition with Fermi surface disappearance without breaking global U(1) symmetry or translational symmetry? 
Our attention is brought to an earlier work\cite{Adams:2011kb}, where we consider a toy model of bosonic system under asymptotic NRCFT, 
where a U(1) symmetry is broken by condensed boson fields around Schr\"odinger black hole.
Bosonic quantum phase transition is likely found there at low temperature phase as \figref{bQPT}. 
It is unavoidable to ask whether these two phase transitions in \figref{fQPT}, \figref{bQPT} have any similar nature.
On one side, $\beta > \beta^*$ of \figref{fQPT} shows a conducting phase with Fermi surface with unbroken U(1) symmetry;
$\Omega > \Omega_*$ side of \figref{bQPT} shows a metallic state with unbroken U(1) symmetry.
On the other side, $\beta < \beta^*$ of \figref{fQPT} shows Fermi surface disappearance;
$\Omega < \Omega_*$ side of \figref{bQPT} shows a superfluid state with broken U(1) symmetry.
Though the two systems have similar asymptotic NRCFT background, one should be aware that the two systems are rather different. 
The bulk side of fermionic model has a charged Schr\"odinger black hole, where the gauge field is chosen to be fixed, the Dirac fermion is in a probe limit. On the other hand, the bulk side of bosonic model in \cite{Adams:2011kb} has a neutral Schr\"odinger black hole, where both the gauge field and bosons are in a probe limit. 
The comparison with a Hawking-Page like transition such as \cite{Hartong:2010ec} would be interesting.\\

\noindent
(3)
It will be illuminating to address more about the phase with Fermi surface disappearance.
In the phase without Fermi surface, there is no superconducting gap opened up in the spectral function.
Specifically we do not introduce any pairing term(such as Yukawa spinor-scalar pairing) in the bulk action, 
so it is not a superconducting phase\footnote{ 
A set-up along \cite{Chen:2009pt, Faulkner:2009am} is the next-step toward boson-fermion interaction, with a superconducting state under NRCFT background.}. 
We only can suspect that tuning $\beta$ from large to small effectively implies tuning fermion interaction from weaker coupling to stronger coupling
- from more Fermi-liquid-like($z\simeq1$) to non-Fermi liquids($z>1$) to strongly correlation 
smear the $A(k,\omega)$ discontinuity into continuity near $\omega_F,k_F$.
Whether one can understand more about the nature of this Fermi surface disappearance, we leave this for future study.\\

\noindent
(4) It is of considerable interest to perform rigorous holographic 
renormalization for spinors to justify holographic dictionary of Green's function, 
following \cite{Leigh:2009ck,Guica:2010sw,vanRees:2012cw}. 
The issue of the proposed counterterms, being totally local\cite{Leigh:2009ck} or non-local\cite{Guica:2010sw}, has not found complete agreement in the literature.
We remark that the discrepancy in holographic renormalization seems to persist and remain to be satisfactorily resolved.\\

%
On the other hand, a subtle issue is that the conformal dimensions $\Delta$ of Schr\"odinger spinors have 
$\nu_{\pm}$ with peculiar $m \pm 1/2$ dependence, distinct from AdS case\cite{Liu:2009dm,Iqbal:2009fd}. 
In AdS case, 
there are
two sets of two component spinors(in \cite{Liu:2009dm,Iqbal:2009fd} notation, $D$ and $A$ for the standard quantization, $B$ and $C$ for the alternate quantization).
In Schr\"odinger case, there are doubled sectors, i.e. 
$\nu_+$ sectors (as $\mathbf{S_1}$,$\mathbf{R_1}$) and $\nu_-$ sectors (as $\mathbf{S_2}$,$\mathbf{R_2}$) shown in Eq.\ref{UV1},\ref{UV2}. 
There is only one independent parameter left for each projected spinor in each sector.
One may wonder why the Green's function in Schr\"odinger case does not possess the two-component spinor structure as in AdS case?
We emphasize that Green's function with two-component spinor structure (such as \cite{Akhavan:2009ns}'s result) has two problems. First, it is  known that the (two-component) spinor structure does not appear in a free non-relativistic fermion theory (analogous to non-relativistic bosons) as discussed in \cite{Leigh:2009ck}.  At this level, \cite{Leigh:2009ck} and our work find an agreement - there is no apparent gamma matrices/spinor structure in the final two-point fermionic Green's function. 
The second problem is that, we find that this approach will sacrifice the distinction between the standard quantization and the alternate quantization, 
which is unreasonable. 
These two known issues seem to suggest our dictionary is a sensible approach.\\

While we may not have the final word in the correct prescription, our results show very interesting physical features,  in particular the numerical results for the Fermi frequency matches closely an analytic guess based on physical reasoning. This should be another piece of supporting evidence that we are capturing the correct physics. \\

\noindent
(5)
Our model is 
a $2+1$D fermionic system. It will be important to study a system in $3+1$D which may exhibit fermions at unitarity\cite{Son:2008ye, Nishida:2010tm}. 
It will also be interesting to explore dynamical exponents other than $z=2$. Indeed gravity duals of finite density systems with asymptotic Schr\"odinger isometry for $d\neq2$, $z\neq2$ are known\cite{Herzog:2008wg,Kovtun:2008qy,Wang:2013}.\\

\noindent
(6) The Schr\"odinger black hole at zero temperature has finite entropy, which implies that our theory may not describe a unique ground state but an ensemble of low energy states.
Moreover, it has been pointed out that discrete lightcone quantization and $\beta$ deformation from the parent AdS black hole\cite{Adams:2008wt,Herzog:2008wg,Maldacena:2008wh,Balasubramanian:2010uw,Barbon:2009az} causes peculiar free energy scaling $F \sim -T^4/\mu^2$ for the system. It will be interesting to know whether bosons or fermions with a full consideration of spacetime back reaction can change the physics of our study, especially the IR AdS$_2$ geometry. 

\noindent
(7) It will be interesting to explore the electron star\cite{Hartnoll:2009ns,Hartnoll:2010gu,Hartnoll:2010ik,Cubrovic:2011xm}
 in the context of Schr\"odinger asymptotic geometry.

\section*{Acknowledgments}

We would like to thank 
A.~Adams,
G.~Baskaran,
L.Y.~Hung,
N.~Iqbal, 
H.~Liu, 
R.~Mahajan,
J.~McGreevy, 
D.~Mross,
Y.~Nishida, 
K.~Rajagopal,
T.~Senthil,
D.~Vegh,
and
W.~Witczak-Krempa 
for valuable discussions.
This work is supported by the U.S. Department of Energy under cooperative research agreement Contract Number DE-FG02-05ER41360.
Research at Perimeter Institute is supported by the
Government of Canada through Industry Canada and by the Province of Ontario
through the Ministry of Research. 
\appendix

\section{$\AdS_2$ scaling} \label{AdS2}

The near horizon geometry of charged Schr\"odinger black hole can be obtain by taking $r=r_0+\epsilon$, with the coordinate redefinition:\\
\be
\tilde{\tau}=K_0^{-1/3}(\tau+\frac{\beta^2r_0^2}{1+\beta^2 r_0^2} y), \;\; \tilde{y}=\frac{K_0^{1/6}}{\sqrt{12}}y, \;\;  \tilde{x_i}=\frac{K_0^{-1/3}}{\sqrt{12}}x, 
\ee
where $K_0=(1+\beta^2 r_0^2)^{-1}$. The metric is
\be
ds^2=- \epsilon^2 d\tilde{\tau}^2/R^2_{\AdS_2}+R^2_{\AdS_2}(d\epsilon^2/\epsilon^2)+r_0^2 (d\tilde{y}^2+d\tilde{x_1}^2+d\tilde{x_2}^2)/R^2_{\AdS_2},
\ee
 with $R_{\\AdS_2}=\frac{(1+\beta^2 r_0^2)^{1/6}}{\sqrt{12}}R=\frac{K_0^{-1/6}}{\sqrt{12}}R$, which is $\AdS_2\times \mathbb{R}^3$ metric. 
The gauge field near horizon is:
$A=A_\tau d\tau=(A_\tau K_0^{1/3}) d\tilde{\tau}+(-A_\tau \beta^2 r_0^2 K_0^{5/6} \sqrt{12})d\tilde{y}$, with $A_\tau\simeq \frac{2Q}{R^2 r_0^3}\epsilon=\frac{Q K_0^{-1/3}}{6 R_{\AdS_2}^2 r_0^3} \epsilon$.
One can solve the Dirac equation, 
with this AdS$_2$ background and supporting gauge field. 
We take the AdS$_2$ rescaling as in \cite{Imeroni:2009cs}, send $\tau \rightarrow \tau/\lambda$ and $\epsilon \rightarrow \lambda \epsilon$ with $\lambda  \rightarrow 0$. In this case, the Dirac equation near the AdS$_2$ boundary becomes,
\be
(\frac{\epsilon}{R_{\AdS_2}} \Gamma_r \partial_{\epsilon}+\frac{1}{2R_{\AdS_2}} \Gamma_r  +C_{\tilde{\tau}} \Gamma_ \tau+ C_{\tilde{y}} \Gamma_y+C_{\tilde{x_1}} \Gamma_ {x_1}-m)\psi=0
\ee
Rewrite the above in terms of two sets of two-component spinors and ``square'' the operator to make it a second order differential equation, we find the ``AdS$_2$'' scaling dimension is the exponent $\nu$ of $\psi \propto \epsilon^{-{1\over2}\pm \nu}$.
\be
\nu= R_{\AdS_2} \sqrt{m^2+C_{\tilde{\tau}}^2-C_{\tilde{y}^2}-C_{\tilde{x_1}^2} }
\ee
with
$C_{\tilde{\tau}}=-i\frac{qQ}{6r_0^3 R_{\AdS_2}}$, $C_{\tilde{y}}=i R_{\AdS_2} \sqrt{12} K_0^{-1/6}(-\beta \omega+\frac{\ell}{2\beta})$, $C_{\tilde{x_1}} =i R_{\AdS_2} k \sqrt{12} K_0^{1/3}$.\\
 
\section{Numerical Set-Up for Green's function} \label{append:Green}

For the foreseeing convenience, we define a matrix $\Dm$, which satisfies Dirac equation Eq.(\ref{eq-dirac}), $\phi' \equiv  \partial_r \phi= \Dm \phi$, 
\be \label{Der}
\Dm \equiv \frac{1}{r\sqrt{f}}
\left(
\begin{matrix}
m K^{-1/6} & 0 & -u-v & -ik/r \\
0 &  m K^{-1/6} & -ik/r & -u+v\\
-u+v & ik/r & -m K^{-1/6}  & 0 \\
ik/r & -u-v & 0 & -m K^{-1/6} 
\end{matrix}
\right)
\ee

We define a converting matrix ${\Cv}$ as a function of $r$ as
\be \label{cov}
{\Cv} \equiv 
r^{1/2}
\left[
\begin{matrix}
r^{\nu_+-1}(a_{1+}+a_{2+}r^{-2})& 
r^{-\nu_+-1}(\alpha_{1+}+\alpha_{2+}r^{-2})& 
r^{\nu_-}(b_{1+}+b_{2+}r^{-2})& 
r^{-\nu_-}(\beta_{1+}+\beta_{2+}r^{-2})\\
r^{\nu_+-1}(a_{1-}+a_{2-}r^{-2})& 
r^{-\nu_+-1}(\alpha_{1-}+\alpha_{2-}r^{-2})& 
r^{\nu_-}(b_{1-}+b_{2-}r^{-2})& 
r^{-\nu_-}(\beta_{1-}+\beta_{2-}r^{-2})\\
r^{\nu_+}(c_{1+}+c_{2+}r^{-2})& 
r^{-\nu_+}(\gamma_{1+}+\gamma_{2+}r^{-2})& 
r^{\nu_--1}(d_{1+}+d_{2+}r^{-2})& 
r^{-\nu_--1}(\delta_{1+}+\delta_{2+}r^{-2})\\
r^{\nu_+}(c_{1-}+c_{2-}r^{-2})& 
r^{-\nu_+}(\gamma_{1-}+\gamma_{2-}r^{-2})& 
r^{\nu_--1}(d_{1-}+d_{2-}r^{-2})& 
r^{-\nu_--1}(\delta_{1-}+\delta_{2-}r^{-2})
\end{matrix}
\right]
\ee
and a set of functions $\mathbf{S_1}(r), \mathbf{S_2}(r), \mathbf{R_1}(r), \mathbf{R_2}(r)$ can be defined from Eq.(\ref{phi-conv}).
This field-redifinition $\mathbf{S_1}(r)$, $\mathbf{S_2}(r)$, $\mathbf{R_1}(r)$, $\mathbf{R_2}(r)$ goes to $\mathbf{S_1}, \mathbf{S_2}, \mathbf{R_1}, \mathbf{R_2}$ at $r \rightarrow \infty$.\\

The EOM of $\mathbf{G}(r)$ in the bulk gravity is 
$
\label{eom1-G}
\mathbf{G}'(r) = \mathbf {R}'(r)\mathbf{S}(r)^{-1} - \mathbf{G}(r) \mathbf {S}(r)'\mathbf {S}(r)^{-1}
$.
Apply Eq.(\ref{Der}) and Eq.(\ref{phi-conv}), then $\mathbf{S}'(r)$ and $\mathbf{R}'(r)$ can be simplified in terms of linear combination of $\mathbf{S}(r)$ and $\mathbf{R}(r)$
\be
\left[
\begin{matrix}
\mathbf{S_1}'(r)\;
\mathbf{R_1}'(r)\;
\mathbf{S_2}'(r)\;
\mathbf{R_2}'(r)
\end{matrix}
\right]^{\text{T}} =\big( {\Cv}^{-1} \phi(r) \big)'= ({\Cv}^{-1} \Dm {\Cv} - {\Cv}^{-1} {\Cv}')  \left[
\begin{matrix}
\mathbf{S_1}(r)\;
\mathbf{R_1}(r)\;
\mathbf{S_2}(r)\;
\mathbf{R_2}(r)
\end{matrix}
\right]^{\text{T}}
\ee

The matrix $D_g(r) \equiv {\Cv}^{-1} \Dm {\Cv} - {\Cv}^{-1} {\Cv}'$ simplifies EOM to,
\be 
\label{eom2-G}
\mathbf{G}'(r) =\left[ \begin{matrix} D_{g2,1} & D_{g2,3}\\ D_{g4,1}& D_{g4,3} \end{matrix} \right]+\left[ \begin{matrix} D_{g2,2} & D_{g2,4}\\ D_{g4,2}& D_{g4,4} \end{matrix} \right]\cdot\mathbf{G}(r)
-\mathbf{G}(r)\cdot \big( \left[\begin{matrix} D_{g1,1} & D_{g1,3}\\ D_{g3,1}& D_{g3,3} \end{matrix} \right] +\left[\begin{matrix} D_{g2,1} & D_{g2,3}\\ D_{g4,1}& D_{g4,3} \end{matrix} \right] \cdot\mathbf{G}(r) \big)
\ee  

Numerically we solve this bulk EOM Eq.(\ref{eom2-G}) of Green's function with the initial condition
to obtain physical results of Eq.(\ref{Ginfty}).\\

Our program code for numerical computation is shared through this URL\cite{code}.

\section{Holographic dictionary for the alternative quantization} \label{append:alternate}

In 
Sec.\ref{dict}, we set up the Holographic dictionary for the standard quantization.
Here we also walk through the similar set-up for the alternative quantization.
Consider the subleading term of $\phi_-$ contributes as a source field, then
\be
\psi_-=(-g g^{rr})^{-1/4} \phi_-\simeq r^{-2} \phi_- \simeq \mathbf{S_2} D_1 r^{\nu_- -5/2}
\ee
which corresponds to the source $\chi_-$,
\be
\chi_-=\lim_{r \rightarrow \infty} r^{\frac{5}{2}-\nu_{-}} \psi_-\simeq \mathbf{S_2}D_1
\ee
$\chi_-$ is proportional to $\mathbf{S_2}$.
The momentum field $\bar{\Pi}_-$ is
\be
\bar{\Pi}_-=\sqrt{-g g^{rr}}  {\psi}_+ = {(-g g^{rr})}^{1/4}  {\phi}_+  \simeq r^2 {\phi}_+ \simeq \mathbf{S_2} B_1 r^{\nu_- +5/2} +\mathbf{R_2} \beta_1 r^{-\nu_- +5/2}+\dots
\ee
which corresponds to the response $\mathcal{O_-}$, 
\be
\mathcal {\bar{O}_-}=\lim_{r \rightarrow \infty} r^{\nu_{-}-\frac{5}{2}} \bar{\Pi}_- \simeq \mathbf{R_2} \beta_1 
\ee
$\mathcal{O_-}$ is proportional to $\mathbf{R_2}$. On the other hand, we can go through the same logic again, though consider the subleading term of $\phi_+$ contributes as a source field, then
\be
\psi_+=(-g g^{rr})^{-1/4} \phi_+\simeq r^{-2} \phi_+ \simeq \mathbf{S_1} A_1 r^{\nu_+ -5/2}
\ee
which corresponds to the source $\chi_+$,
\be
\chi_+=\lim_{r \rightarrow \infty} r^{\frac{5}{2}-\nu_{+}} \psi_+\simeq \mathbf{S_1}A_1
\ee
$\chi_+$ is proportional to $\mathbf{S_1}$.
The momentum field $\bar{\Pi}_+$ is
\be
\bar{\Pi}_+=-\sqrt{-g g^{rr}} {\psi}_- =- {(-g g^{rr})}^{1/4} {\phi}_-  \simeq - r^2{\phi}_- \simeq -\mathbf{S_1} C_1 r^{\nu_+ +5/2} - \mathbf{R_1} \gamma_1 r^{-\nu_+ +5/2}+\dots
\ee
which corresponds to the response $\mathcal{O_+}$, 
\be
\mathcal{O_+}=-\lim_{r \rightarrow \infty} r^{\nu_{+}-\frac{5}{2}} \bar{\Pi}_+ \simeq \mathbf{R_1} \gamma_1 
\ee
$\mathcal{O_+}$ is proportional to $\mathbf{R_1}$. Now we again derive $\mathbf{S_1},\mathbf{S_2}$ are identified as sources, $\mathbf{R_1},\mathbf{R_2}$ are identified as responses for this alternative quantization.

\section{Pure Schr\"odinger Green's function at zero T zero density} \label{append:pureSchr}

\subsection{two-point correlators at the leading order}

Here we compare the pure Schr\"odinger two-points function at zero T zero density of Ref.\cite{Akhavan:2009ns} with our formulation\footnote{In \cite{Akhavan:2009ns}, their $\epsilon$ coordinates are inverse of our $r$, also their $d=3$ is our $d=2$ case}. Specifically, in $d=2$, they show, 
\be
\langle \psi_M(x,t) \bar{\psi}_M(0,0)  \rangle \propto r^{2\nu_+}
\ee
Here we crosscheck our analysis indeed matches theirs. The boundary action is $\int_{\partial \mathcal{M}} dt d\xi d^2x \sqrt{-g g^{rr}} \bar{\psi} \psi$, with $ \sqrt{-g g^{rr}}=r^4$, by plugging our UV boundary expansion in Eq.(\ref{UV1}), Eq.(\ref{UV2}), with $\psi=(-g g^{rr})^{-1/4} \phi \simeq r^{-2} \phi$, we arrive $\bar{\psi} \psi|_{\partial \mathcal{M}} \propto r^{-4} \bar{\phi} \phi $. Notice $\Gamma_{\tau}$ in $\bar{\phi} \phi $ coupling the 1st to the 3rd component of $\phi$, meanwhile coupling the 2nd to the 4th component of $\phi$. With leading piece in $\phi$ is the 4th component of $\phi$, which is $\mathbf{S_1}  \, r^{\nu_++\frac{1}{2}}$. This couples to the leading order of the 2nd component of $\phi$, which is $\mathbf{S_1} \, r^{\nu_+-\frac{1}{2}} a_2$. 
Neglect other factors and coefficients,
\be
\int_{\partial \mathcal{M}} dt d\xi d^2x \sqrt{-g g^{rr}} \bar{\psi} \psi \propto \bar{\phi} \phi \propto r^{2\nu_+}
\ee
One can obtain the scaling form of two-point correlator $\langle \psi_M(x,t) \bar{\psi}_M(0,0) \rangle $ at the leading order $r^{2\nu_+}$.
The scaling $r^{2\nu_+}$ matches for Ref.\cite{Akhavan:2009ns},\cite{Leigh:2009ck} and our works, however the precise form of our correlator is {\it not} identical with Ref.\cite{Akhavan:2009ns}.
We should note that both Ref.\cite{Leigh:2009ck} and our work contain higher order terms in the fermion field source/response, thus both works contain $r^{2\nu_+}$, $r^{2\nu_-}$ scaling, 
while Ref.\cite{Akhavan:2009ns} only contains $r^{2\nu_+}$ scaling. In the next subsection, we will delve further into our Green's function and its similarity with that of Ref.\cite{Leigh:2009ck}.

\subsection{Pure Schr\"odinger Green's function from response over source}
Solve the Dirac's equation in zero T zero density Schr\"odinger spacetime $ds^2=-r^4dt^2+2r^2dt d\xi+r^2 d\vec{x}^2+dr^2/r^2$, the bulk fields have the following form: 
\begin{align}
\psi_+&=&r^{-\frac{d+3}{2}}K_{\nu_+}(\mathbf{k}/ r)V^+&+&g_+(\mathbf{k},r) \Gamma_\xi U^+ &+& r^{-\frac{d+3}{2}}K_{-\nu_+}(\mathbf{k}/ r) V^- &+&g_-(\mathbf{k},r) \Gamma_\xi U^-  \\
\psi_-&=&f_+(\mathbf{k},r) \Gamma_\xi V^+&+&r^{-\frac{d+3}{2}}K_{\nu_-}(\mathbf{k}/ r) U^+ &+& f_-(\mathbf{k},r) \Gamma_\xi V^- &+&r^{-\frac{d+3}{2}}K_{-\nu_-}(\mathbf{k}/ r) U^-
\end{align}

where $d$ is the spatial dimension of $\vec{x}$, the quantity 
\be
\mathbf{k}=\sqrt{-2\ell \omega+k^2} \nonumber
\ee
 is a coordinate-invariant form of momentum. 
We denote $U^+,V^+$ as two components spinor bases for $\nu_\pm$ series(the first two columns), and denote $U^-,V^-$ as two components spinor bases for $-\nu_\pm$ series(the last two columns),  
$K_{\nu}(\mathbf{k}/ r)$ is modified Bessel function, solution of 
$r^2\partial_r^2 K_{\nu}(\mathbf{k}/ r)+r\partial_r K_{\nu}(\mathbf{k}/r)-\big( (\mathbf{k}/r)^2+\nu^2)K_{\nu}(\mathbf{k}/r\big)=0$. The expansion of $K_{\nu_{\pm}}(\mathbf{k}/r), f_{\pm}(\mathbf{k},r), g_{\pm}(\mathbf{k},r),\psi_+,\psi_-$ near $r \rightarrow \infty$ are
\be
K_{\nu_{\pm}}(\mathbf{k}/r) = 2^{-1+\nu_\pm} (\mathbf{k}/r)^{-\nu_\pm} \Gamma(\nu_\pm) (1+ O(r^{-2}))
\ee
\bea
f_{\pm}(\mathbf{k},r)=i\frac{2^{\pm\nu_{+}-2}\ell\; \Gamma(\pm \nu_+)}{\pm\nu_++m+\frac{1}{2}} k^{\mp \nu_+} r^{-\frac{d+1}{2} \pm\nu_+} (1+ O(r^{-2}))\\
g_{\pm}(\mathbf{k},r)=-i\frac{2^{\pm\nu_{-}-2}\ell\; \Gamma(\pm \nu_-)}{\pm\nu_--m+\frac{1}{2}} k^{\mp \nu_-} r^{-\frac{d+1}{2} \pm\nu_-}  (1+ O(r^{-2}))
\eea
\begin{align}
\label{exp1}
\psi_+&=&A& r^{\nu_+-\frac{d+3}{2}}(1+O(r^{-2}))+B r^{\nu_--\frac{d+1}{2}}(1+O(r^{-2})) +\alpha r^{-\nu_+-\frac{d+3}{2}}(1+O(r^{-2}))+\beta r^{-\nu_--\frac{d+1}{2}}(1+O(r^{-2})) \\
\label{exp2}
\psi_-&=&C&r^{\nu_+-\frac{d+1}{2}}(1+O(r^{-2}))+D r^{\nu_--\frac{d+3}{2}}(1+O(r^{-2})) + \gamma r^{-\nu_+-\frac{d+1}{2}}(1+O(r^{-2}))+\delta r^{-\nu_--\frac{d+3}{2}}(1+O(r^{-2}))
\end{align}

Notice $\psi_+=(-gg_{rr})^{-{1\over4}} \phi_+$ and $\psi_-=(-gg_{rr})^{-{1\over4}} \phi_-$, we can compare this expansion respect to Eq.(\ref{UV1}), Eq.(\ref{UV2}). The expansion matches, with the projection constrains on the spinors:
\bea 
\label{proj1}
C=\frac{-\ell}{2\beta(\nu_++m+\frac{1}{2})}P_-A, \;\;\; B=\frac{-\ell}{2\beta(\nu_- -m+\frac{1}{2})}P_+D,\\
\label{proj2}
\gamma=\frac{-\ell}{2\beta(-\nu_++m+\frac{1}{2})}P_- \alpha, \;\;\; \beta=\frac{-\ell}{2\beta(-\nu_- -m+\frac{1}{2})}P_+\delta,
\eea
In the case of the charged Schr\"odinger black hole for Eq.(\ref{UV1}), Eq.(\ref{UV2}), the projection relation Eq.(\ref{proj1}), Eq.(\ref{proj2})'s $\ell$ is replaced by $\ell+qQ\beta$ for  Eq.(\ref{project1}), Eq.(\ref{project2}).

There are extra constraints on two-component spinors $V^{\pm},\; U^{\pm}$:
\be
V^{\pm}=\frac{-i}{2 \ell}(i k^\mu \Gamma_\mu) \Gamma_\xi V^{\pm}, \;\;\; U^{\pm}=\frac{-i}{2 \ell}(i k^\mu \Gamma_\mu) \Gamma_\xi U^{\pm}
\ee
or equivalently,
\be
\Gamma_\xi V^{\pm}=\frac{2 \ell (k^\mu \Gamma_\mu)}{\mathbf{k}^2} V^{\pm}, \;\;\; \Gamma_\xi U^{\pm}=\frac{2 \ell (k^\mu \Gamma_\mu)}{\mathbf{k}^2} U^{\pm}
\ee

where $k^\mu \Gamma_\mu= \ell \Gamma_t - \omega \Gamma_\xi+ k_x \Gamma_x$.
By identifying source and response based on our holographic dictionary, we have the response and source matrix,\\
\begin{align}
\mathbf{R}&=&\begin{bmatrix}  
\mathbf{R}_\mathbf{1}^1&\mathbf{R}_\mathbf{1}^2\\
\mathbf{R}_\mathbf{2}^1&\mathbf{R}_\mathbf{2}^2
\end{bmatrix} 
&=&\begin{bmatrix} 
2^{-\nu_+-1} \Gamma(-\nu_+) \mathbf{k}^{\nu_+} (V^-)_2^{(1,2)}\\
2^{-\nu_--1} \Gamma(-\nu_-) \mathbf{k}^{\nu_-} (U^-)_3^{(1,2)}\\
\end{bmatrix}&=\begin{bmatrix} 
2^{-\nu_+-2} \Gamma(-\nu_+) \mathbf{k}^{\nu_+} \frac{-i}{\ell} (i k^\mu \Gamma_\mu \Gamma_\xi V^-)_2^{(1,2)}\\
2^{-\nu_--2} \Gamma(-\nu_-) \mathbf{k}^{\nu_-}  \frac{-i}{\ell} (i k^\mu \Gamma_\mu  \Gamma_\xi U^-)_3^{(1,2)}\\
\end{bmatrix}\\
\mathbf{S}&=&\begin{bmatrix} 
\mathbf{S}_\mathbf{1}^1&\mathbf{S}_\mathbf{1}^2\\
\mathbf{S}_\mathbf{2}^1&\mathbf{S}_\mathbf{2}^2
\end{bmatrix}
&=&\begin{bmatrix} 
i \frac{2^{\nu_+-2} \ell \Gamma(\nu_+)}{\nu_++m+\frac{1}{2}} \mathbf{k}^{-\nu_+} ( \Gamma_\xi V^+)_4^{(1,2)}\\
-i \frac{2^{\nu_--2} \ell \Gamma(\nu_-)}{\nu_--m+\frac{1}{2}} \mathbf{k}^{-\nu_-} ( \Gamma_\xi  U^+)_1^{(1,2)}\\
\end{bmatrix}&
\end{align}

Here we follow the notation in Sec.\ref{sec3-IR}, introducing upperindices $(1,2)$ to distinguish the first and the second sets of two independent 
boundary conditions for spinors.
We also introduce lower indices $j=1,2,3,4$, implying the $j$-th component of 4-spinor.
For example, $(V^+)^{(2)}_4$ means reading the $4$-th component of the spinor $(V^+)$ from the second$(2)$ type of 
boundary condition.


For the notation convenience, we define,
\bea
\mathbf{V^-}^{(1,2)}=(k^\mu \Gamma_\mu\Gamma_\xi V^-)^{(1,2)}_2&,& \;\;\; \mathbf{U^-}^{(1,2)}=(k^\mu \Gamma_\mu\Gamma_\xi U^-)^{(1,2)}_3\\
\mathbf{V^+}^{(1,2)}=(\Gamma_\xi V^+)^{(1,2)}_4&,& \;\;\; \mathbf{U^+}^{(1,2)}=(\Gamma_\xi U^+)^{(1,2)}_1
\eea
Again, the upper indices (1,2) are chocies for the first or the second independent 
boundary conditions. 
The lower indices $1,2,3,4$ are indices for spinor components.  
Green's function is,

\be
\mathbf{G}=\begin{bmatrix} 
\mathbf{G}_\mathbf{1}^1&\mathbf{G}_\mathbf{1}^2\\
\mathbf{G}_\mathbf{2}^1&\mathbf{G}_\mathbf{2}^2
\end{bmatrix}
\ee

with each component
\bea \label{G4comp}
\mathbf{G}_\mathbf{1}^1= \frac{-i}{\ell^2}(\frac{\mathbf{k}}{2})^{2 \nu_+}\frac{\Gamma(-\nu_+)}{\Gamma(\nu_+)}(\nu_++m+\frac{1}{2})  \frac{  (  \mathbf{V^-}^{(1)} \mathbf{U^+}^{(2)} -\mathbf{V^-}^{(2)} \mathbf{U^+}^{(1)} ) }{( \mathbf{V^+}^{(1)} \mathbf{U^+}^{(2)}-\mathbf{V^+}^{(2)} \mathbf{U^+}^{(1)} )}\\
\mathbf{G}_\mathbf{1}^2= \frac{i}{\ell^2} (\frac{\mathbf{k}}{2})^{ \nu_++\nu_-}\frac{\Gamma(-\nu_+)}{\Gamma(\nu_-)}(\nu_- -m+\frac{1}{2})  \frac{(  -\mathbf{V^-}^{(1)} \mathbf{V^+}^{(2)} +\mathbf{V^-}^{(2)} \mathbf{V^+}^{(1)} )}{ ( \mathbf{V^+}^{(1)} \mathbf{U^+}^{(2)}-\mathbf{V^+}^{(2)} \mathbf{U^+}^{(1)} )} \\
\mathbf{G}_\mathbf{2}^1= \frac{-i}{\ell^2}  (\frac{\mathbf{k}}{2})^{ \nu_++\nu_-}\frac{\Gamma(-\nu_-)}{\Gamma(\nu_+)}(\nu_++m+\frac{1}{2})   \frac{ \mathbf{U^-}^{(1)} \mathbf{U^+}^{(2)} -\mathbf{U^-}^{(2)} \mathbf{U^+}^{(1)}  }{( \mathbf{V^+}^{(1)} \mathbf{U^+}^{(2)}-\mathbf{V^+}^{(2)} \mathbf{U^+}^{(1)} )}\\
\mathbf{G}_\mathbf{2}^2= \frac{i}{\ell^2} (\frac{\mathbf{k}}{2})^{ 2\nu_-}\frac{\Gamma(-\nu_-)}{\Gamma(\nu_-)}(\nu_- -m+\frac{1}{2})  \frac{(  -\mathbf{U^-}^{(1)} \mathbf{V^+}^{(2)} +\mathbf{U^-}^{(2)} \mathbf{V^+}^{(1)}    )}{ ( \mathbf{V^+}^{(1)} \mathbf{U^+}^{(2)}-\mathbf{V^+}^{(2)} \mathbf{U^+}^{(1)} )} 
\eea

We know that Green's function in \cite{Akhavan:2009ns} contains only the $r^{2\nu_+}$ contribution.
Our two-point Green's function closely resembles that of Ref.\cite{Leigh:2009ck} with subleading structure, $r^{2\nu_+},r^{2\nu_-}$. 
A quick way to check this, is that comparing to Eq (79) of \cite{Leigh:2009ck},
their $\langle \mathcal{O}^\dagger_+ \mathcal{O}_+\rangle \simeq (k^2-2 \ell \omega)^{ \nu_+}$ scales identically as 
$\mathbf{G}_\mathbf{1}^1 \sim \mathbf{k}^{2\nu_+}= (k^2-2 \ell \omega)^{ \nu_+}$ of ours.
And their $\langle \mathcal{O}^\dagger_- \mathcal{O}_-\rangle \simeq (k^2-2 \ell \omega)^{ \nu_-}$ scales identically as 
$\mathbf{G}_\mathbf{2}^2 \sim \mathbf{k}^{2\nu_-}=(k^2-2 \ell \omega)^{ \nu_-}$ of ours. 





\end{document}